\documentclass[reprint,nofootinbib,...]{revtex4-1} 
\usepackage{amsmath}%
\usepackage{amsthm,amssymb}
\usepackage{graphicx}
\usepackage{epstopdf}
\usepackage{xcolor}
\usepackage{algcompatible}
\usepackage[size=small]{caption}
\usepackage{etoolbox}
\usepackage{booktabs}
\usepackage{multirow}
\usepackage[utf8]{inputenc}
\usepackage[colorinlistoftodos]{todonotes}

\usepackage{subfig}

\usepackage[hyperfootnotes=true]{hyperref}
\hypersetup{
 colorlinks=true,
 citecolor=blue,
 linkcolor=blue,
 urlcolor=blue}
 
 \DeclareMathOperator{\sign}{sign}

\newcommand{\nconst}{64}       % number of jet constituents

\newcommand{\JetPtMin}{300}    % minimum jet pT [GeV]
\newcommand{\JetMassMin}{50}   % minimum jet mass [GeV]

\newcommand{\HLLayers}{3}
\newcommand{\HLUnits}{384}
\newcommand{\HLAuc}{0.79}

\newcommand{\PfnPhiLayers}{three}
\newcommand{\PfnFLayers}{three}
\newcommand{\PfnPhiUnits}{256,256,64}
\newcommand{\PfnFUnits}{256}
\newcommand{\PfnAuc}{0.88}

\newcommand{\AdvLayers}{4}
\newcommand{\AdvUnits}{300}

\newcommand{\pt}{p_\mathrm{T}} % needs mathmode

\begin{document}

\title{AI Safety for High Energy Physics}

\author{Benjamin Nachman}
\thanks{Authors contributed equally}

\affiliation{Physics Division, Lawrence Berkeley National Laboratory, Berkeley, CA 94720, USA}
\email{bpnachman@lbl.gov}

\author{Chase Shimmin}
\thanks{Authors contributed equally}

\affiliation{Department of Physics, Yale University, New Haven, CT 06511, USA}
\email{chase.shimmin@yale.edu}

\begin{abstract}

The field of high-energy physics (HEP), along with many scientific disciplines, is currently experiencing a dramatic influx of new methodologies powered by modern machine learning techniques.
Over the last few years, a growing body of HEP literature has focused on identifying promising applications of deep learning in particular, and more recently these techniques are starting to be realized in an increasing number of experimental measurements.
The overall conclusion from this impressive and extensive set of studies is that rarer and more complex physics signatures can be identified with the new set of powerful tools from deep learning.  However, there is an unstudied systematic risk associated with combining the traditional HEP workflow and deep learning with high-dimensional data.
In particular, calibrating and validating the response of deep neural networks is in general not experimentally feasible, and therefore current methods may be biased in ways that are not covered by current uncertainty estimates.
By borrowing ideas from AI safety, we illustrate these potential issues and propose a method to bound the size of unaccounted for uncertainty.
In addition to providing a pragmatic diagnostic, this work will hopefully begin a dialogue within the community about the robust application of deep learning to experimental analyses.
\end{abstract}

\date{\today}
\maketitle

%%%%%%%%%%%%%%%%%%%%%%%%%%%%%

\section{Introduction}
\label{sec:intro}
Experiments in collider-based high-energy physics (HEP) rely critically on detailed simulations which model length scales from sub-nuclear reactions all the way to macroscopic detector-length scales in order to connect fundamental theories to experimentally-observable quantities.
Typical experiments, such as measurements of physical constants or searches for new particle species, are designed using blinded methodology, and depend on these calibrated simulations to predict the relative rates of background and signal events.
These predictions are in turn used to define the statistical significance and/or confidence intervals of the results observed after unblinding.
While the simulations involved in this process are highly sophisticated, they are only an approximation to reality and therefore systematic mismodeling must be accounted for by calibrating to data, when possible, and by assessing systematic uncertainties.

Traditionally, partitions of the data known as \textit{signal} and \textit{control regions} are defined by applying selective criteria on physical observables, in order to isolate regions of the data that are expected to be sensitive to the phenomena of interest from well-understood phenomena.
It is then possible to validate and/or calibrate the simulated background predictions against data observed in the signal-free region without biasing the blinded analysis.
However, it is often the case that several different observables, perhaps following some complicated relationship, are useful for defining such regions of the data.
A typical application of machine learning in HEP is to automate the construction of signal and control regions by reformulating the task as an optimization problem; for example, a binary classifier may be trained on simulations to label observed events as signal-like or background-like.
While this has been done for years using ``shallow'' classifiers such as Boosted Decision Trees, the success of these methods have generally depended strongly on the choice of features input to the classifier, incurring a significant amount of effort towards ``feature engineering'' to identify useful \textit{high-level} observables.

With the more recent introduction of deep learning methods, it has become possible to construct increasingly elaborate classification models using higher-dimensionality input features.
Perhaps surprisingly, it has been shown\footnote{There are now too many examples to cite them all here.  References~\cite{Larkoski:2017jix,Radovic:2018dip,Guest:2018yhq} are recent reviews of deep learning in HEP and References~\cite{Baldi:2014kfa,deOliveira:2015xxd,Baldi:2016fql,Guest:2016iqz} were the earliest applications of deep learning to collider-based HEP classification problems.} that when provided with high-dimensional \textit{low-level} (HDLL) features (\textit{i.e.} observables that are minimally processed using physical intuition), deep neural networks are able to automatically learn to exceed the performance of networks trained on physically-motivated high-level features.

Increasingly, experimentalists at the Large Hadron Collider (LHC) and elsewhere are taking this message seriously.
While the first analysis-level deep learning results from the LHC are only starting to become public (see e.g.~\cite{Aad:2019yxi,ATLAS-CONF-2019-017,CMS-PAS-SUS-19-009}), analysis-non-specific deep learning models have been used for a few years, notably with early successful applications in flavor tagging~\cite{CMS-DP-2017-005,ATL-PHYS-PUB-2017-003}.
In addition, there is a plethora of experimental and phenomenological studies for additional methods which will likely be realized as part of physics analyses in the near future.
This includes proposals to use the lowest-level inputs available from the detector, reaching input feature dimensionalities of $\mathcal{O}(10^5)$~\cite{Andrews:2018nwy,Andrews:2019faz} and beyond.

A potential problem with this approach arises when combining deep learning on HDLL features with the conventional simulation-based analysis paradigm.  In the traditional approach, uncertainties on the extrapolation between the control region and the signal region rely on simulation variations that can be validated on a small number of one-dimensional physically-motivated features.  Correlated uncertainties covering the full HDLL feature space are often not known or experimentally infeasible~\cite{Nachman:2019dol}.
Moreover, low-dimensional validation may be insufficient: recent developments in the area of Artificial Intelligence (AI) safety have demonstrated that when neural networks operate on high-dimensional input spaces, classification performance of neural networks can be arbitrarily degraded by applying subtle variations to the input features~\cite{Szegedy14intriguingproperties,DBLP:journals/corr/GoodfellowSS14}.

To illustrate this challenge, we implement an adversarial attack that demonstrates small perturbations in detector-level measurements can have drastic effects on the performance of a neural network trained to identify signal-like events.
This adversarial procedure can be conceptualized as a sort of \textit{demon}~\cite{TheLordKelvin} which intercedes between a physical process and the observation of that process, in such a way as to maximally confound a neural network while remaining minimally noticeable by current experimental standards.
While certainly no such demon exists, we propose a procedure based on this concept as a diagnostic tool to evaluate the worst-case sensitivity of a deep network-based anaysis to uncertainty from mismodeling high-dimensional correlations.
This bound may then be used either to demonstrate that a given network architecture is manifestly robust against a certain class of systematic uncertainties, or as a guiding metric to aid in the future development of more robust networks.

\section{Benchmark Problem}
\label{sec:benchmark}

As one of the first examples from HEP in which both feature engineering and deep learning have demonstrated promising advantages, jet classification~\cite{Larkoski:2017jix} is a natural testing ground for this study.
Jets are collimated sprays of particles resulting from high energy quark and gluon fragmentation, which are clustered~\cite{Cacciari:2008gp} into groups that approximate physical states in the original hard scattering process.
A typical problem is to identify whether a jet originated from a quark/gluon state, or by the decay of some intermediate massive particle.
In particular, we simulate two $pp$ scattering processes: a background consisting of dijets, and a signal comprised of $Z$-bosons produced in association with an energetic photon, in which the Lorentz-boosted $Z$ particle subsequently decays to a pair of quarks.

All samples are simulated at parton-level using Madgraph5~\cite{Alwall:2014hca}, with fragmentation by Pythia~8~\cite{Sjostrand:2006za,Sjostrand:2007gs}, and  an ATLAS-like detector simulation by Delphes 3.4.1~\cite{deFavereau:2013fsa}.
Activated regions of the simulated calorimeter detector (`towers') are clustered using the anti-$k_t$ algorithm with radius parameter $R=1$~\cite{Cacciari:2008gp}, and only the highest\footnote{Using collider coordinates, $\pt$ is the particle momentum transverse to the collision axis, $\phi$ is the azimuthal angle and the pseudo-rapidity $\eta$ is $-\ln(\tan(\theta/2))$, where $\theta$ is the polar angle.}-$\pt$ jet of each event is considered for the classification task.
Furthermore, selected events must contain a jet with $\pt>\JetPtMin$ GeV and mass $m>\JetMassMin$ GeV/$c^2$.
This jet must also be comprised of at least three constituent towers.

After clustering, a jet $J_i$ is represented as a truncated list\footnote{The architecture described below can accommodate a variable number of inputs.  However, the adversarial setup is currently configured to output a fixed size perturbation.  The loss in performance when using more than 64 constituents was negligible.} of $N_i\leq\nconst$ constituent 4-momenta: $J_i = \{(\pt^k, \eta^k,\phi^k) \colon k = 1,...,N_i\}$.  The resulting dimensionality of observable features for learning is thus about 200.
The $\pt$ of each constituent is expressed in units of TeV, resulting in a maximum value that is of order unity, suitable for input to neural networks.
Jets are then input to two different network architectures which are trained to discriminate signal jets from background jets.

The first architecture, referred to as the \textit{High-level} (HL) model, has a first layer with no learnable parameters which computes four jet-level observables from the input constituents: $\pt$, $\eta$, invariant mass, and $D_2^{(\beta=2)}$.
The first three features simply represent the 4-vector associated with the entire jet; the quantity $D_2$~\cite{Larkoski:2014gra} is theoretically motivated from the strong force and designed to identify jets with radiation patterns consistent with two sub-jet axes, characteristic of Lorentz boosted boson decays.
The remainder of the network is comprised of $\HLLayers$ fully-connected layers, each with $\HLUnits$ units and ReLU activation, followed by a single output neuron with sigmoid activation.

The second architecture, referred to as the \textit{Low-level} (LL) model, is a Particle Flow Network~\cite{Komiske:2018cqr}, also with a special non-parameteric first layer.
This layer shifts the $(\eta,\phi)$ values of constituents such that the jet axis is centered at zero.
This common preprocessing step is done within the network so that adversarial attacks are unable to alter the jet origin.
The benchmark LL model has \PfnPhiLayers~layers in the $\Phi$ subnetwork with $\PfnPhiUnits$ units and \PfnFLayers layers in the $F$ subnetwork, each with $\PfnFUnits$ units.
Both subnetworks use ReLU activations on all layers.
The final layer is again a single neuron with sigmoid activation.
We note that similar results were obtained using a simple fully-connected network.
However, the PFN architecture is invariant with respect to permutations of the input constituents, which simplifies the interpretation as reordering caused by the adversary has no impact on the result.  We suspect that any other similarly performing LL network (see e.g. Ref.~\cite{Kasieczka:2019dbj}) will have a similar susceptibility as the PFN.

Both networks are implemented using Keras~\cite{keras} and Tensorflow~\cite{tensorflow}, and are trained using the Adam~\cite{adam} optimizer.
The loss function is the binary cross-entropy to classify signal and background events.
After tuning the architectures via hyperparameter scans, we found the LL network (AUC=$\PfnAuc$) was able to significantly outperform the HL network (AUC=$\HLAuc$), as is often the case.

\section{Methods}
To demonstrate the potential sensitivity of HDLL networks to subtle mismodeling of their input features, we subject each of our benchmark networks to an \textit{adversarial attack}.
An adversarial attack exploits the gradient of the (fixed) target network with respect to its inputs, in order to shift those inputs to solicit the desired response from the classifier.

Note that in order to realize the specific mismodeling necessary to foil a particular classifier, the parameters of the target network must be known; hence the ``demon'' posited in Sec.~\ref{sec:intro}.
Nonetheless, as this attack yields a mathematically optimal perturbation to a given input for a given network, it can be viewed as a worst-case scenario.
Therefore, if a specific network is shown to be robust against the attack, it is also reasonable to conclude that the effects of any intractable systematic mismodelings present in physics simulations are safe to ignore.
Conversely, if a network is shown to be sensitive to such attacks, the adversarially-induced systematic shift can be viewed only as a (potentially weak) upper bound for more realistic systematic effects.
In this case, a poor upper bound may indicate more careful scrutiny of a network's systematic exposure is warranted.
The bound may also be used as a guiding metric in the development of more robust models, as discussed in Sec.~\ref{sec:discussion}.

In this work we implement two different forms of adversarial attack.
The first is based on the fast gradient sign method (FGSM) proposed in Ref.~\cite{DBLP:journals/corr/GoodfellowSS14}, which computes a bounded perturbation for a given input.
The second is a broader attack utilizing an adversarial neural network which learns to construct malicious jets for arbitrary inputs.
The former method is the most literal realization of our ``demon'', as each jet is individually modified to optimally confound the network.
The latter demonstrates that there exists a universal mapping between one dataset (e.g. simulation) and another (e.g. experimental data) which systematically affects network performance.
Therefore the adversary can be thought of as transfer function representing a measurement effect or theoretical mismodeling.

In both cases, our demon is tasked with transforming signal jets to induce a background-like response from the benchmark classifier.
In such a scenario, an experimental analysis optimized using the simulated signal model would be very likely to reject true signal events as background.
This in turn could lead to overconfidently excluding a theoretical hypothesis, or even missing a discovery of new physics altogether.

\subsection{Fast Gradient Sign Method}
\label{subsec:fgsm}
The FGSM method works by taking an individual jet $J_i$ and regressing the loss function to compute a \textit{bounded} perturbation $\delta J_i$, such that network's response to the input $J_i + \delta J_i$ tends towards some desired value.
Specifically, the perturbation is given by:
\begin{equation}
\label{eq:FGSM}
\delta J_i = \left.\sign\left[\nabla_{J} \mathcal{L}_\text{XE}\left(f(J), y_\mathrm{bg}\right)\right]\right\rvert_{J=J_i} \,,
\end{equation}
where $f$ is the target classifier network, $\mathcal{L}_\text{XE}$ is the binary crossentropy loss function, and $y_\mathrm{bg}$ is the label corresponding to background events.  The size of the gradient is scaled to a free parameter $\epsilon$, as described below.  This ensures that the perturbation of each input observable is bounded while moving approximately in the direction normal to the decision boundary. 

In our example, this gradient, like the input, is a $\nconst\times3$ tensor, where the quantities on the second axis represent $\pt$, $\eta$, and $\phi$.
We want to ensure that the perturbation is small relative to experimental resolution, yet these observables have rather different scales associated to them.
To accommodate this, and ensure a physically relevant perturbation, the perturbation is given by $J_i\mapsto J_i(1+\epsilon_{\pt}\delta J_i)+\epsilon_\Omega\delta J_i$, where $\epsilon_{\pt}=(\epsilon,0,0)$ and $\epsilon_\Omega=(0,\epsilon,\epsilon)$.  In principle, these $\epsilon$ do not need to be the same, but are chosen here to all be $\epsilon=0.001$.   All multiplications in Eq.~\ref{eq:FGSM} follow tensorflow broadcast semantics~\cite{tensorflow}.  This procedure is iteratively applied 10 times so the scale of the perturbations are bounded by 0.01.

\subsection{Adversarial Network Method}
\label{sec:adversarial}

In this approach, given a target classifier $f$, we train a second neural network $g$.
The goal is to learn a map $g(J)\mapsto J'$, for an arbitrary signal or background jet $J$ such that the classifier network $f$ will perceive $J'$ as background.
Because various observables $\mathcal{O}$ can be validated against data for backgrounds, it is important that the distribution before $\Pr(\mathcal{O}(J) | J\in\mathrm{bg})$ match the distribution $\Pr(\mathcal{O}(g(J))|J\in\mathrm{bg})$ after perturbation. For practical reasons, we enforce this by preventing the adversary from making large changes to these observables on a jet-by-jet basis for background events.

To further simplify training, we also bound the degree to which $g$ can modify constituents, and prevent it from creating new constituents within a jet.
In particular, the adversary is unable to induce collinear splittings or add spurious soft radiation that physically-motivated observables are often designed to be robust against.  Given these constraints in the form of $g$, the attack presented here represents a worst-case scenario only for a specific class of mismodeling.
As it turns out, even this restricted form of attack can have surprisingly large effects; we leave the assessment of sensitivity to more general attack models to future work.

The adversarial network is trained by minimizing separate loss functions for signal and background defined by:
\begin{align}
\mathcal{L}_\text{sig}&=\log(1-f(g(J))),\\\nonumber
\mathcal{L}_\text{bg}&=\lambda_\mathrm{cls} (f(J)-f(g(J)))^2\\
&\hspace{3mm}+\sum_i \lambda^{(i)}_\mathrm{obs} (\mathcal{O}^{(i)}(J)-\mathcal{O}^{(i)}(g(J))^2\,.
\end{align}
$\mathcal{L}_\mathrm{sig}$ is the categorical crossentropy, which impels $g$ to modify signal jets so as to be labeled as background by $f$.
The first term of $\mathcal{L}_\mathrm{bg}$ minimizes changes between the target network's response to the jet before and after the adversarial perturbation.
The functions $\mathcal{O}^{(i)}(J):\mathbb{R}^{3N}\rightarrow \mathbb{R}$ represent any features of interest to be preserved.
The tunable hyperparameters $\lambda_\mathrm{cls}, \lambda^{(i)}_\mathrm{obs} \geq 0$ encode the adversary's preference to preserve the target network response and observable features, respectively, for background events.

In our experiments, $g$ is a fully-connected network with \AdvLayers\ hidden layers, each with \AdvUnits\ units and ReLU activation.
The penultimate layer has $\nconst\times 3$ units, with $\tanh$ activation.
Analogously to the $\sign$ function in Eq.~\ref{eq:FGSM} and the bounding parameters $\epsilon$ in Sec.~\ref{subsec:fgsm}, the outputs of the final layer are bounded by applying a $\tanh$ activation, and the axes corresponding to $\pt$, $\eta$, and $\phi$ are scaled by parameters $\rho_{\pt}$, $\rho_\eta$, and $\rho_\phi$, respectively.
The output of this layer represents a differential change in the input jet, $\delta J$.
The final layer is essentially a residual skip-connection layer computing $J+\delta J$ as described in Sec.~\ref{subsec:fgsm}.

\begin{figure}[h!]
\centering
\includegraphics[width=0.45\textwidth]{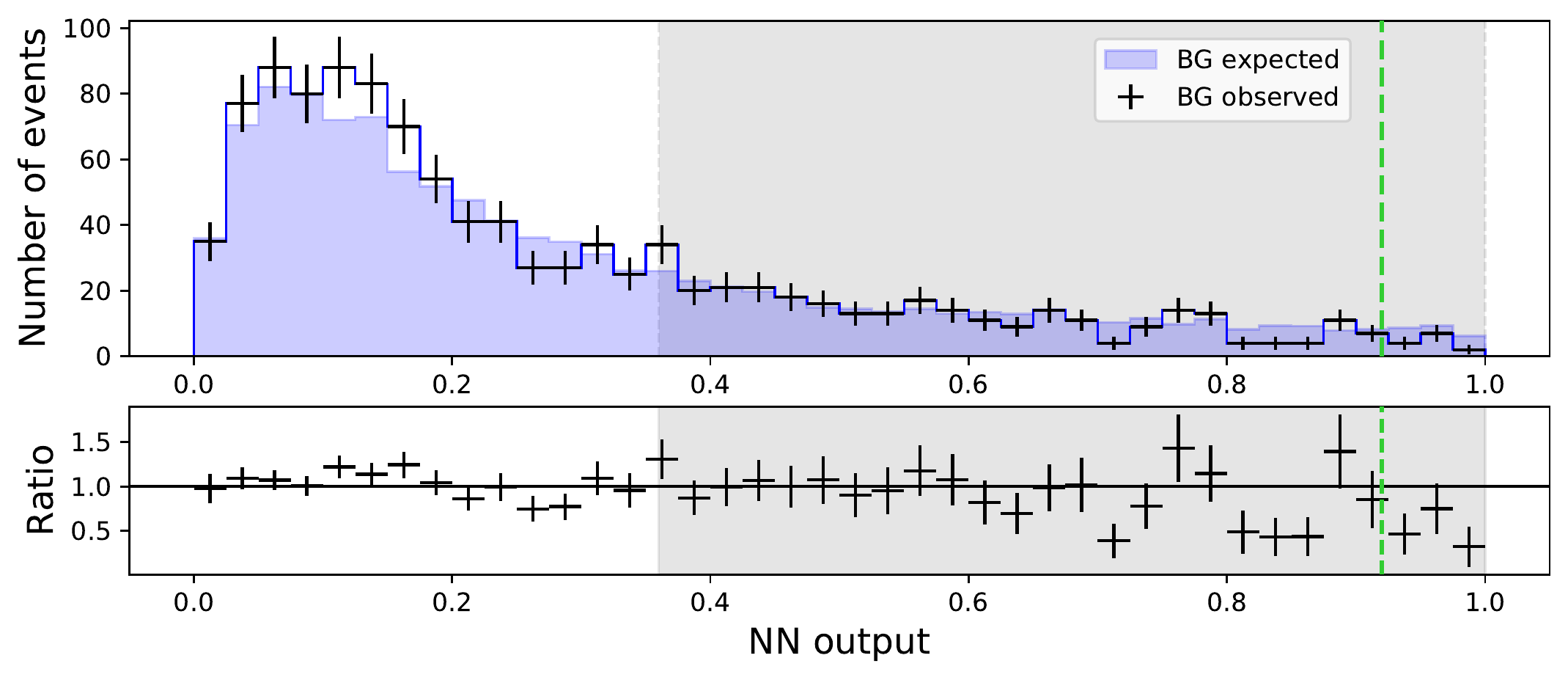} \\
\includegraphics[width=0.45\textwidth]{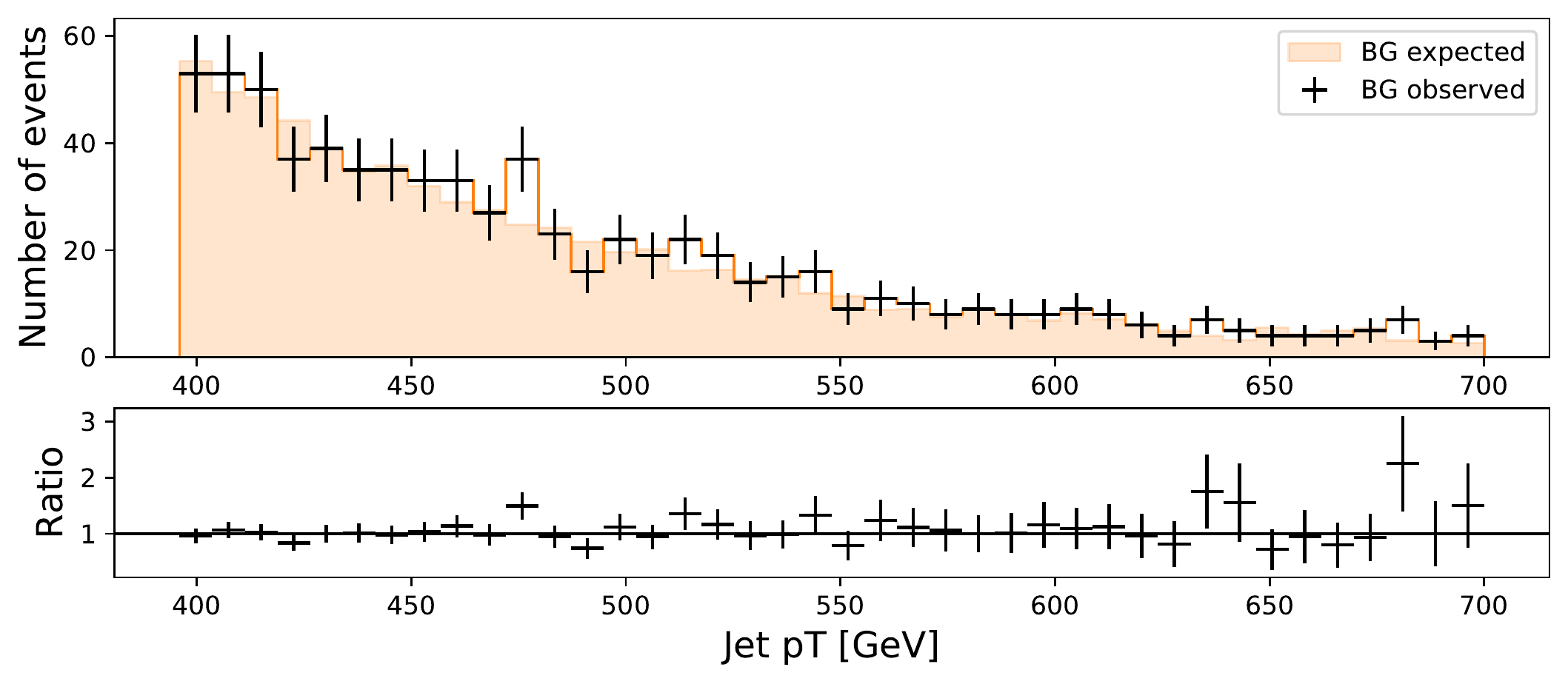} \\
\includegraphics[width=0.45\textwidth]{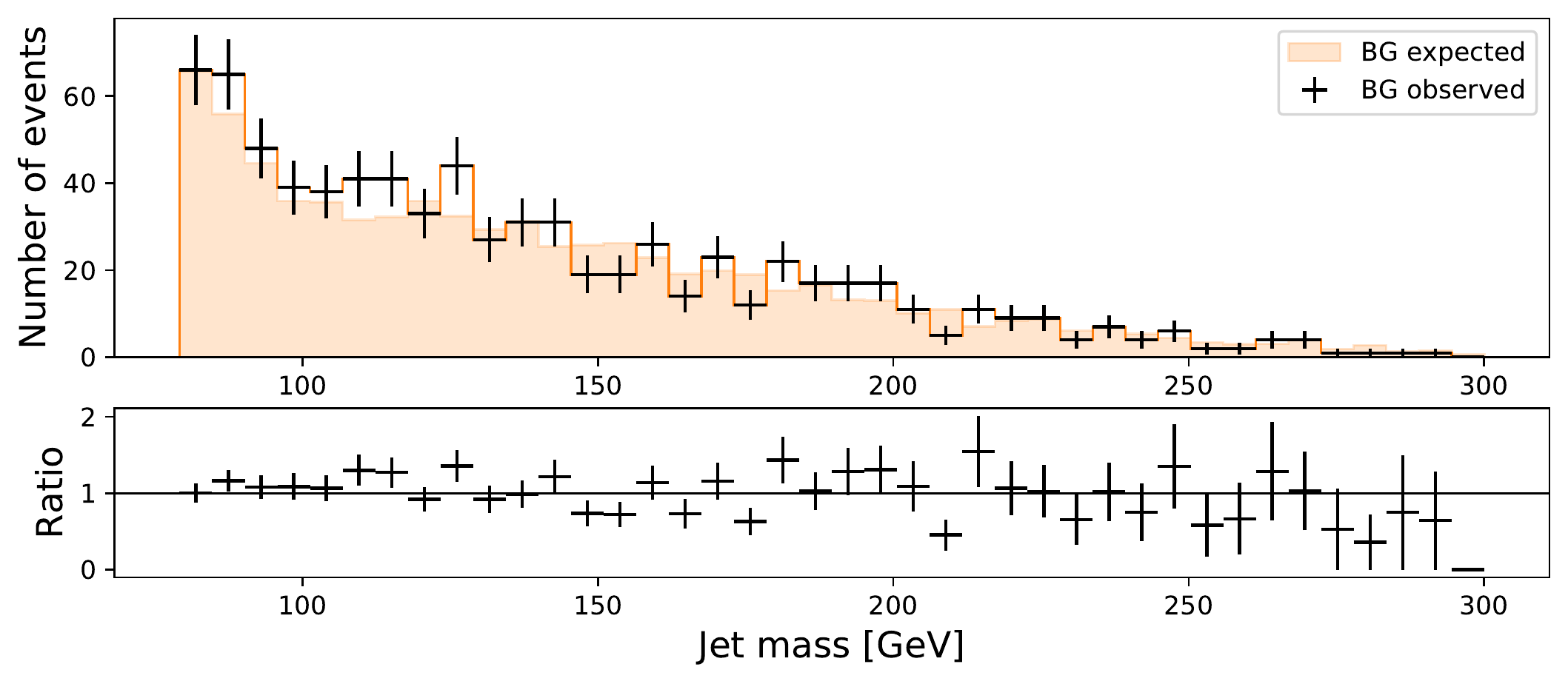} 
\caption{Illustration of typical validation procedure.
Pseudodata (black points) are sampled from the BG distribution with the adversarial perturbation applied; solid histograms show the unperturbed BG model.
Top: The unshaded control region in this case is defined where the signal efficiency is expected to be less than 10\%; the shaded region would typically be blinded when designing an experiment.
The green vertical line indicates the expected optimal signal region.
Middle, Bottom: The jet $\pt$ and mass distributions for events in the control region.
Good agreement is observed between the ``observed'' pseudodata and the expected background model in the control region for all three observables.
The $\chi^2/\mathrm{ndf}$ values are $14.7/14$, $25.0/40$, and $37.8/40$ repsectively.
}
\label{fig:AdvValidation}
\end{figure}

A separate adversary is trained for each of the HL and LL benchmark networks.
In both cases, the bounding magnitude of the constituent perturbations are fixed at $\vec{\rho} = 0.02$, which is slightly larger than the scale of perturbations for the FGSM.
Two observable constraints are included in $\mathcal{L}_\mathrm{bg}$: the jet mass and $\pt$.
The parameters $\lambda_\mathrm{cls}$ and $\lambda_\mathrm{obs}$ are tuned by training until either convergence or until certain validation criteria are violated.
The validation criteria are met when the Kolmogorov-Smirnoff (KS) test statistic between perturbed and unperturbed background distributions are below heuristically-defined thresholds of 0.04 for jet mass and $\pt$, and 0.02 for classifier response.  In practice, these thresholds would be set by the data statistics as well as the size of known experimental uncertainties. 
A more realistic test in practice is to consider the $\chi^2$ agreement between validation histograms evaluated in an unblinded control region, as illustrated in Fig.~\ref{fig:AdvValidation} for the case of the LL network.

\section{Results}
\label{sec:results}

\begin{figure}[h!]
\centering
\includegraphics[width=0.45\textwidth]{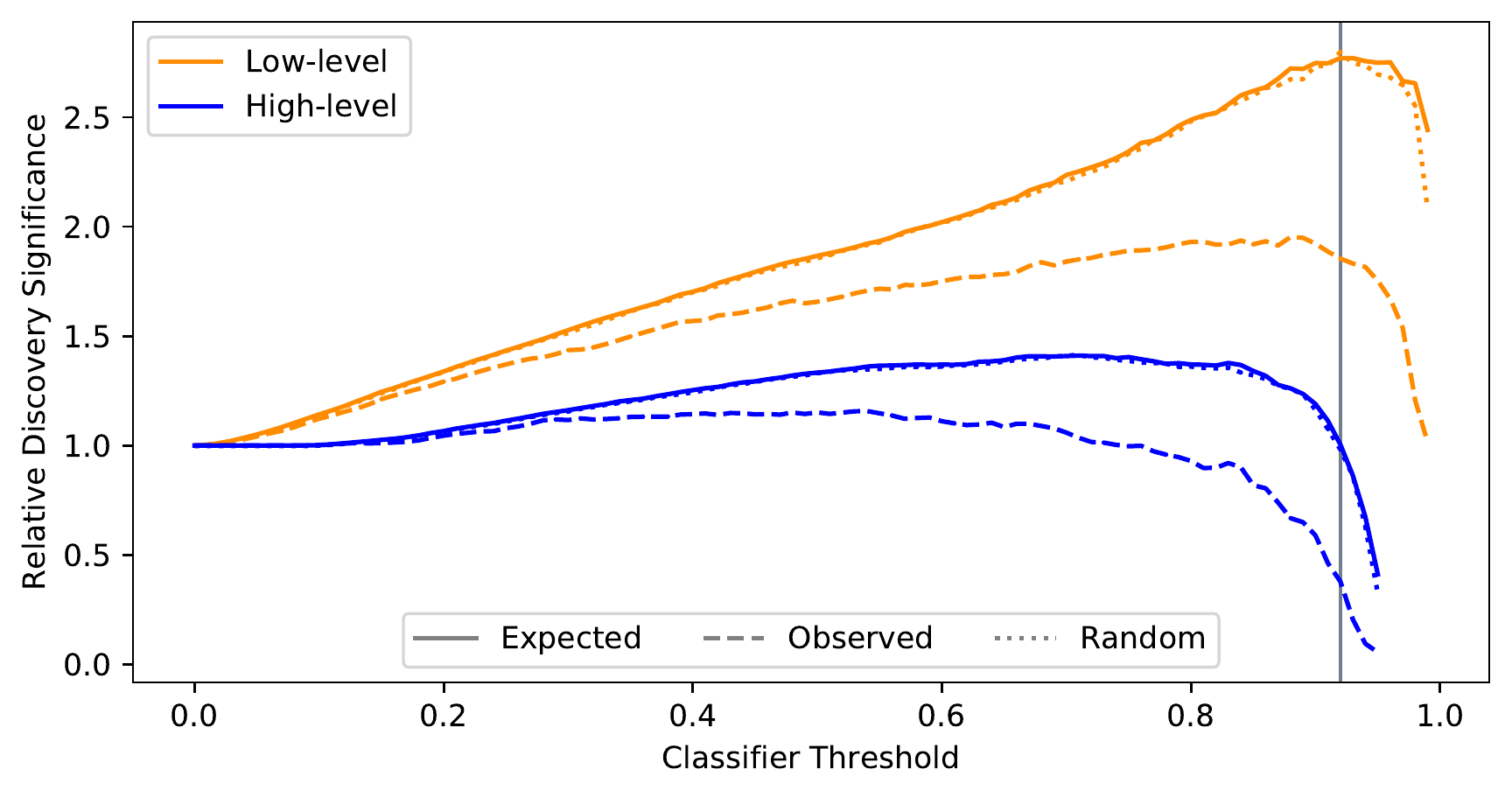}
\caption{The relative discovery significance as a function of the classifier threshold for HL observables and LL observables before and after the FGSM perturbation.  Also shown is the effect induced by randomly perturbing constituents with the same $\epsilon$ values used for the FGSM.}
\label{fig:FGSM2}
\end{figure}

To quantify the effect of these adversarial attacks, we consider a simplified example of a typical experimental analysis in HEP.
If $S$ and $B$ are the predicted number of signal and background events, respectively, then in the asymptotic limit ($S + B\gg 1$, $S\ll B$~\cite{Cowan:2010js}), the expected statistical significance of an observation with respect to the background-only hypothesis is $S/\sqrt{B}$, in units of standard deviations.
After considering only events that pass a classifier threshold, the relative change in the significance is $\epsilon_S/\sqrt{\epsilon_B}$, where $\epsilon_S$ is the true positive rate (signal efficiency) and $\epsilon_B$ is the false positive rate (background efficiency).
A classifier is only useful for improving the sensitivity of a search if this \textit{relative discovery significance} exceeds unity.
The relative discovery significance for both the LL and HL classifiers are shown in Fig.~\ref{fig:FGSM2}.
As expected given it's AUC (Sec.~\ref{sec:benchmark}), the LL classifier yields a more sensitive result than the HL classifier, with peak relative discovery significances of about 2.5 and 1.5, respectively.
Figure~\ref{fig:FGSM2} additionally shows the relative discovery significance after the application of the FGSM.
This bounded perturbation is only applied to signal events and designed to make them look more like background events.
Both the LL and HL relative discovery significance are degraded by this perturbation by about 30\%.
Additionally, the optimal classifier threshold shifts for the HL case, so a threshold chosen based on the nominal simulation would actually have a relative discovery significance less than unity in the perturbed simulation.

\begin{figure}[h!]
\centering
\includegraphics[width=0.35\textwidth]{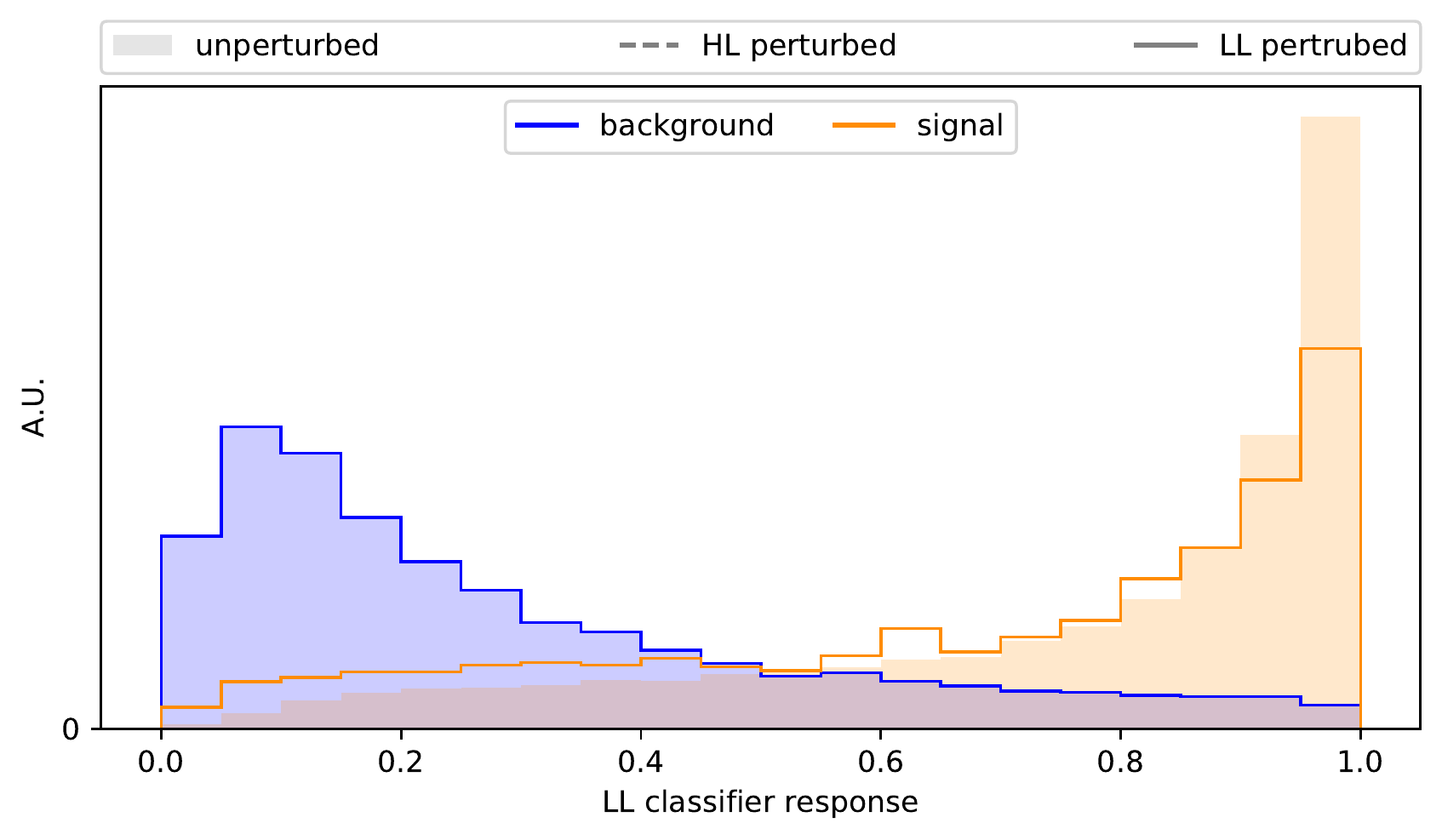}
\includegraphics[width=0.35\textwidth]{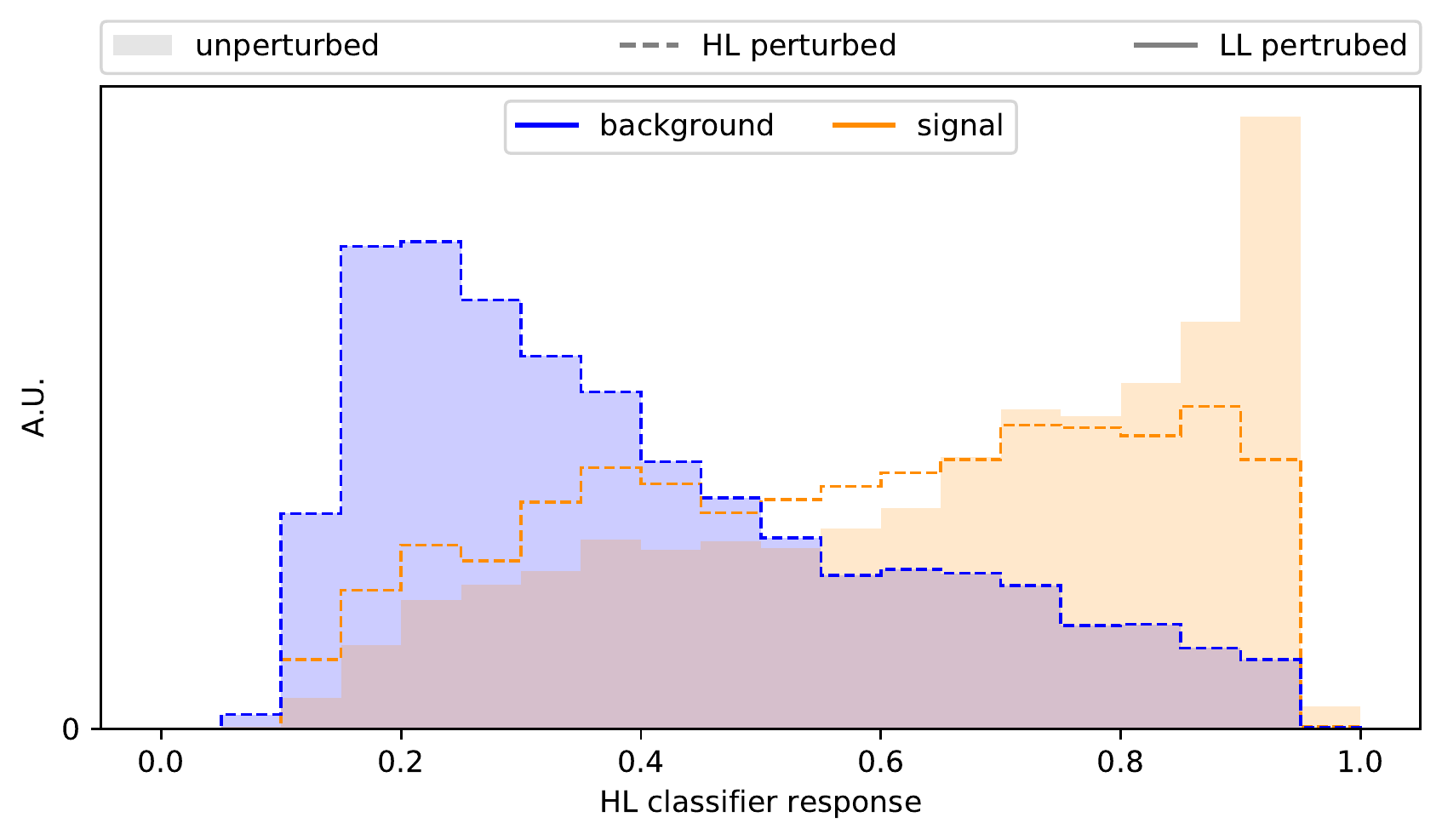}
\includegraphics[width=0.35\textwidth]{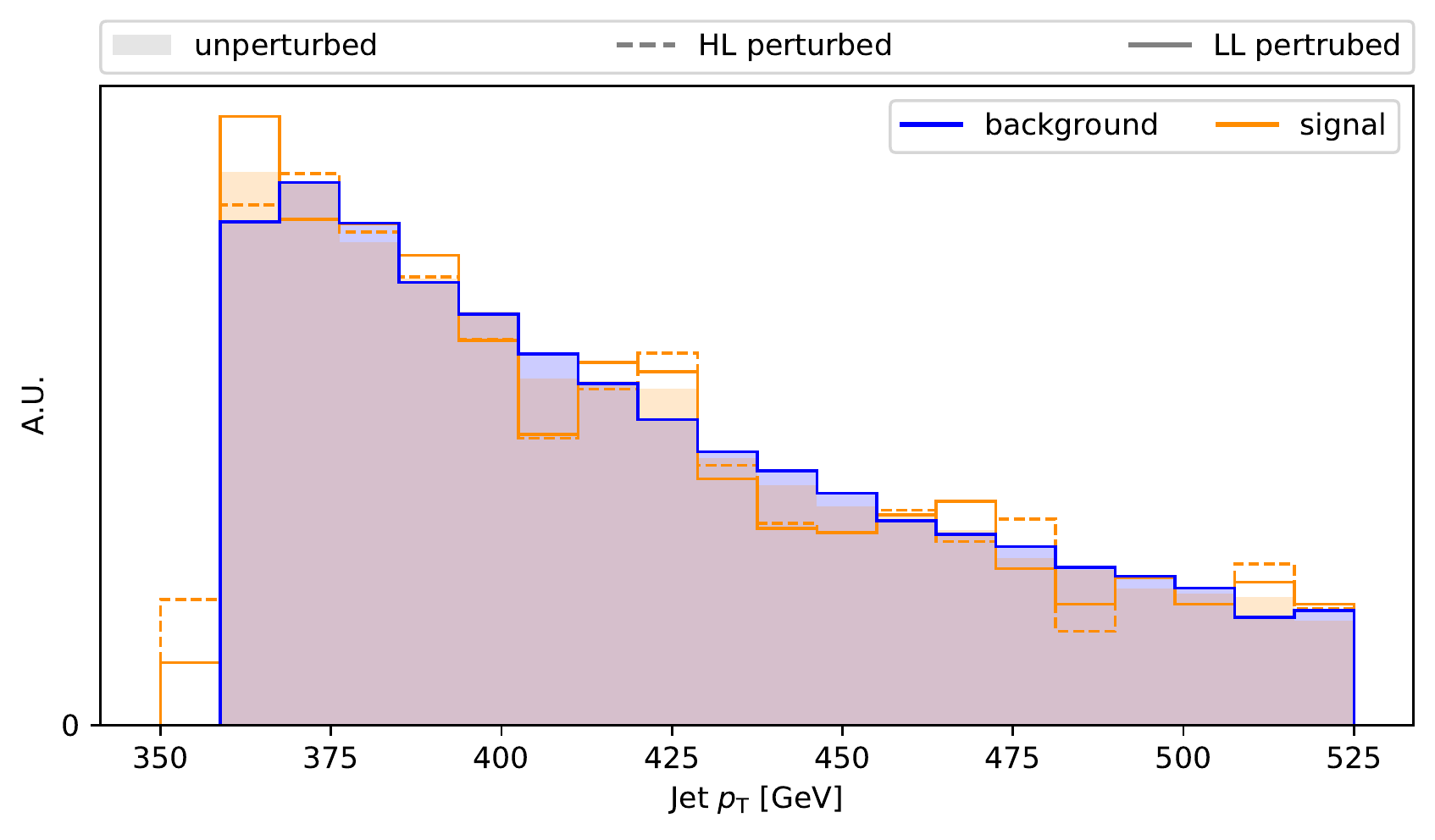}
\includegraphics[width=0.35\textwidth]{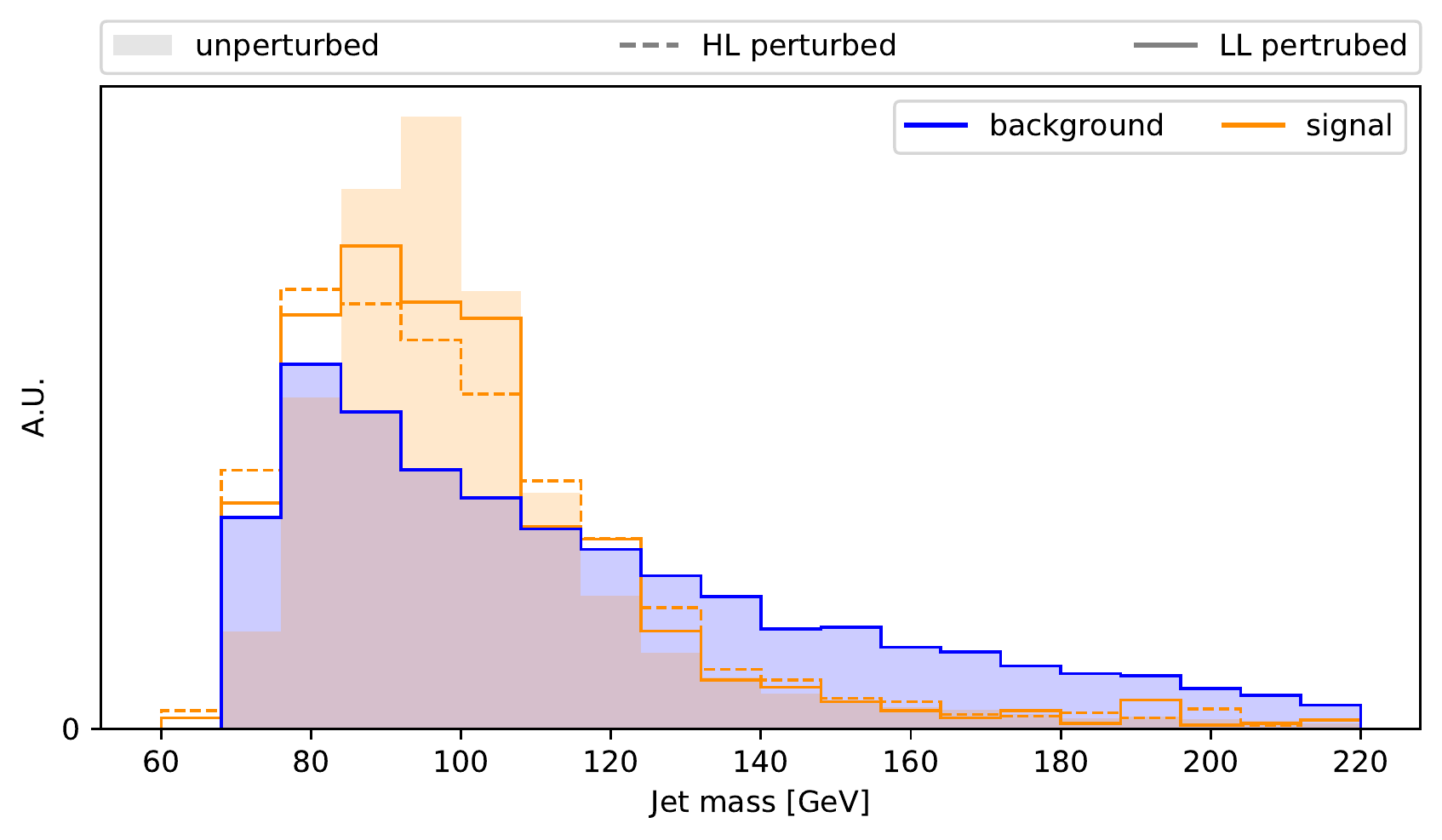}
\includegraphics[width=0.35\textwidth]{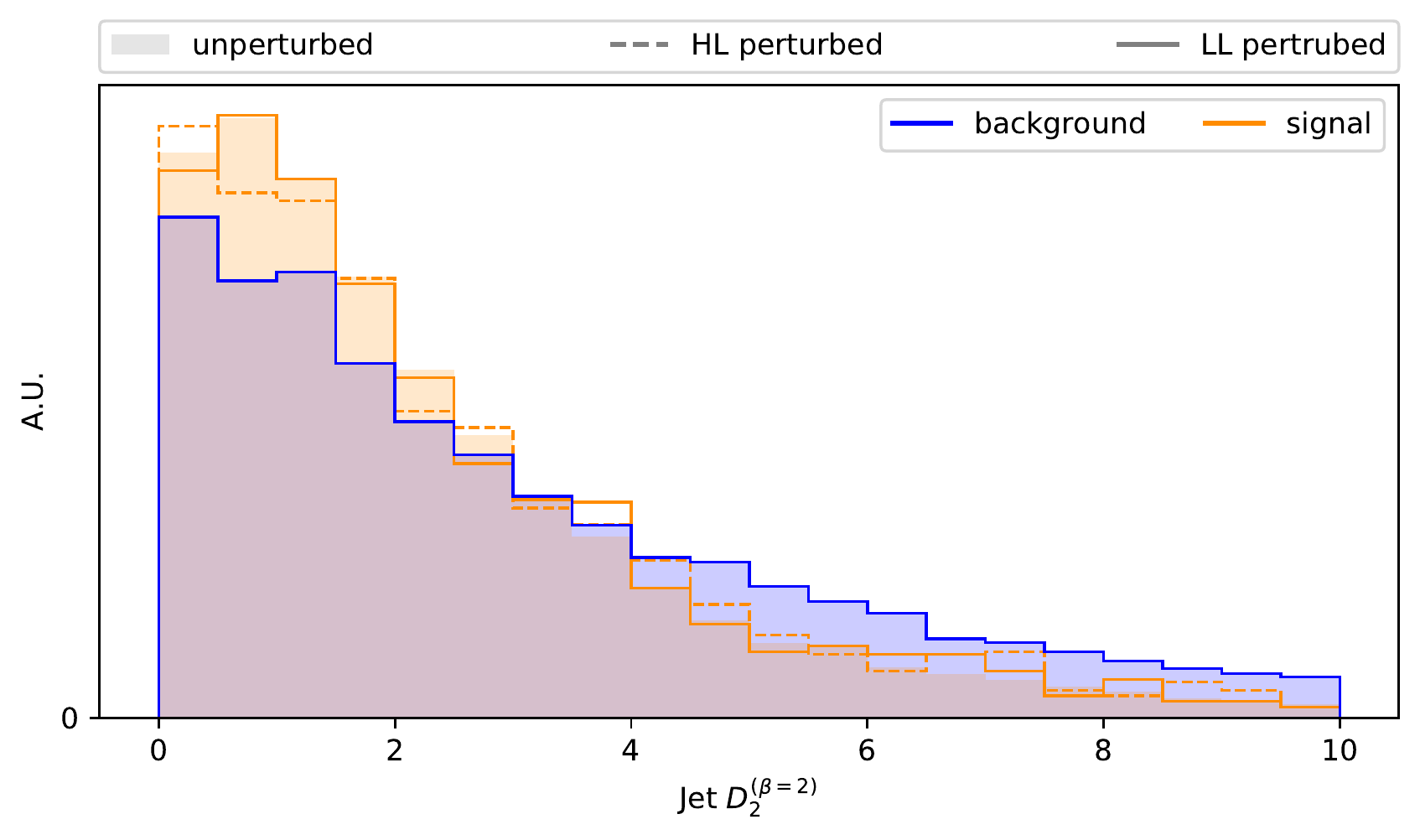}
\caption{The classifier output and various observables before and after the FGSM attack for signal and background.  Only 1500 signal jets are used for the FGSM perturbation.}
\label{fig:FGSM:distributions}
\end{figure}

The FGSM attack has bounded perturbations on the jet constituent four-vectors, but is otherwise unconstrained.
Figure~\ref{fig:FGSM:distributions} shows the modifications to various signal observables as a result of the FGSM attack.
Due to the limited size of the perturbation, the qualitative shapes of the signal $\pt$, mass, and $D_2^{(\beta=2)}$ distributions are the same and after the FGSM perturbation.
Of these, the mass distribution is most affected.
The distributions of the classifier outputs are shifted to the left, overlapping more with the background distribution, hence the degradation observed in Fig.~\ref{fig:FGSM2}.

\begin{figure}[h!]
\centering
\includegraphics[width=0.45\textwidth]{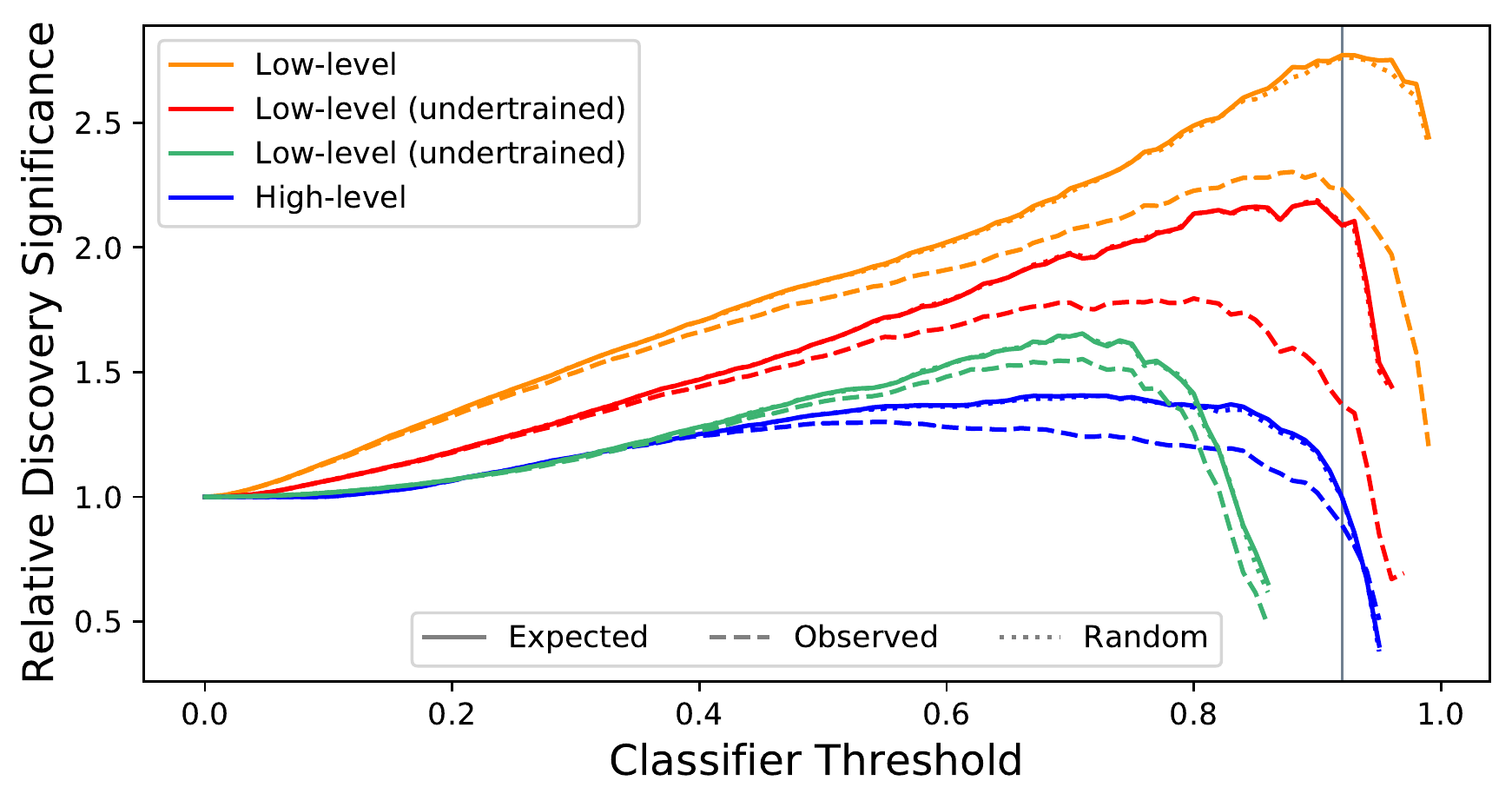}
\caption{Effect of adversarial mismodeling on discovery significance, for high-level and low-level feature networks.
The vertical gray line indicates the expected optimal selection threshold, which differs by about 25\% from the ``true'' significance when taking the adversarial perturbation into account.
The HL network's expected sensitivity differs by about 15\% from the true value.
While the fully-trained low-level network is expected to perform better than the high-level network, it is also more strongly affected by an adversarial attack.
However, when the LL network is deliberately undertrained, its susceptibility is reduced.
Also shown is the effect induced by randomly perturbing constituents by a uniform distribution in the range $[-\rho,+\rho]$.
}
\label{fig:method2b}
\end{figure}

While the FGSM perturbation is illustrative, it is too synthetic because only the signal events are attacked and there are no constraints on HL observables that can be validated.  The adversarial attack described in Sec.~\ref{sec:adversarial} avoids both of these issues, and the analog to Fig.~\ref{fig:FGSM2} is presented in Fig.~\ref{fig:method2b}.
For these discovery significance curves, both the signal and background distributions are perturbed by the same network, representing a consistent mismodeling between the jets expected by the simulation model (pre-perturbation), and the jets actually observed in our hypothetical scenario (post-perturbation).
The degradation in the relative discovery significance is comparable to the FGSM for the fully trained LL network, however, we were unable to train an adversary to produce as large an effect on the HL network.
This is to be expected, as the HL network derives much of its classification power from the jet mass observable, which the adversarial network is constrained to minimally change.

Fig.~\ref{fig:method2b} also demonstrates the effect of the adversarial attack on deliberately undertrained instances of the LL network.
We found that the classifier network's training is interruped early on, the susceptibility to the adversarial attack is reduced, and tends to increase with additional training.
In particular, when the LL network is trained only to the same level of performance as the HL network, it is nearly impervious to the adversarial attack.
We hypothesize that the additional information the LL network uses in order to outperform the HL network is more sensitive to small-scale perturbations than the theoretically-motivated HL observables.
Although this effect seems to have spurious counterexamples due to random network initialization, the trend may  suggest that undertraining very sensitive HDLL networks could be be a useful regularization technique to build in analysis robustness, while still providing a performance boost relative to HL architectures.

\begin{figure}[h!]
\centering
\includegraphics[width=0.35\textwidth]{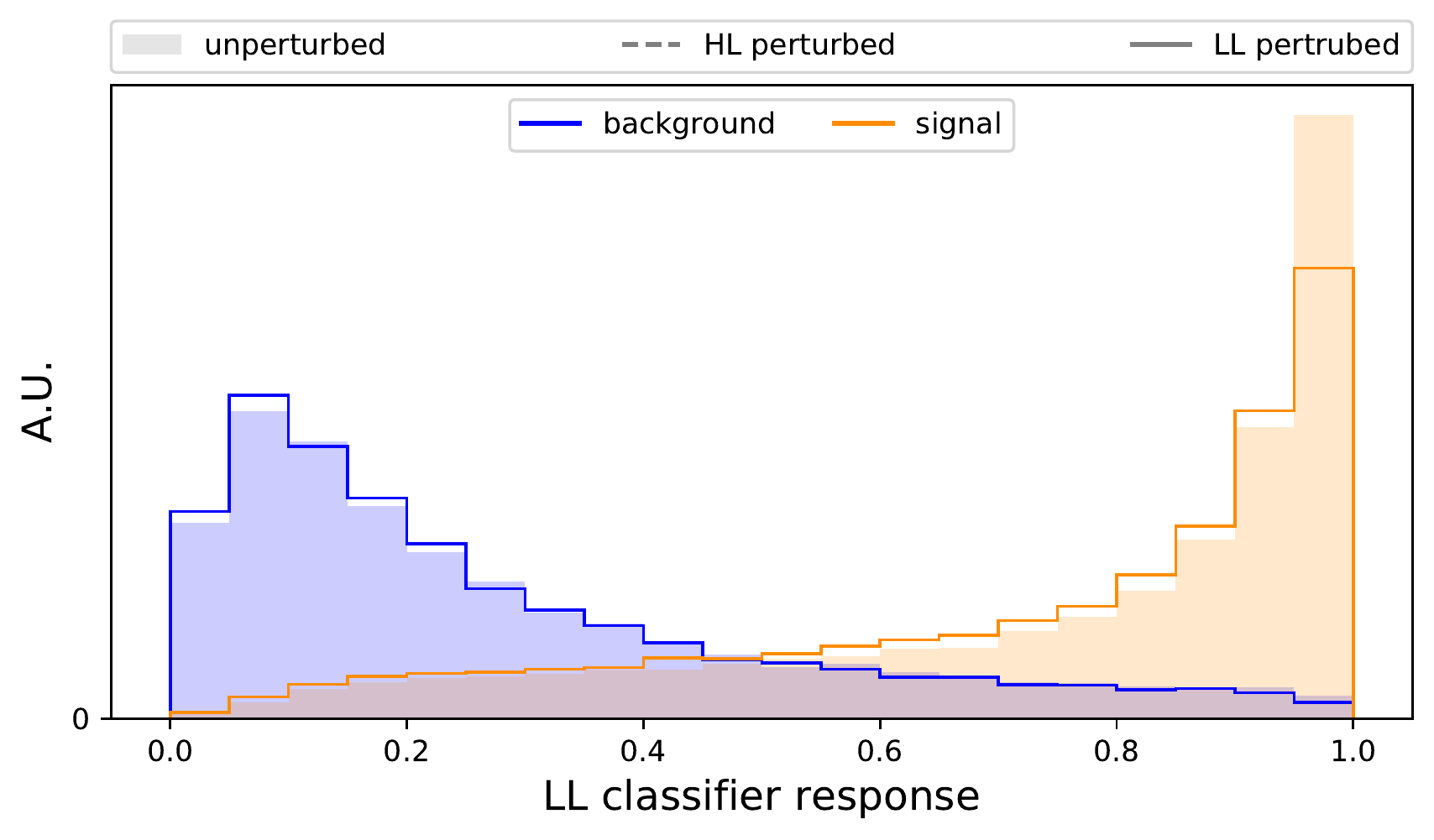} \\
\includegraphics[width=0.35\textwidth]{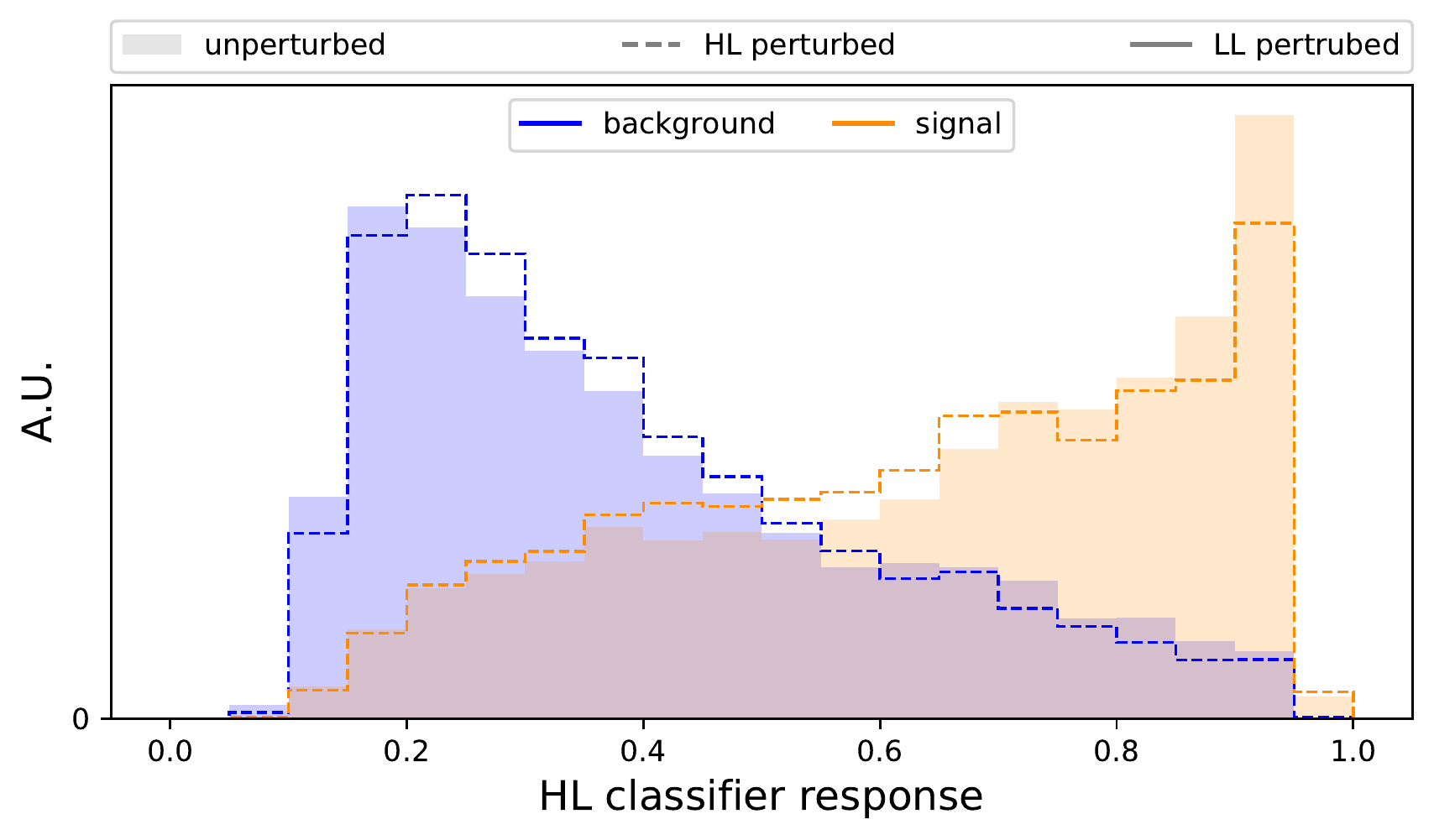} \\
\includegraphics[width=0.35\textwidth]{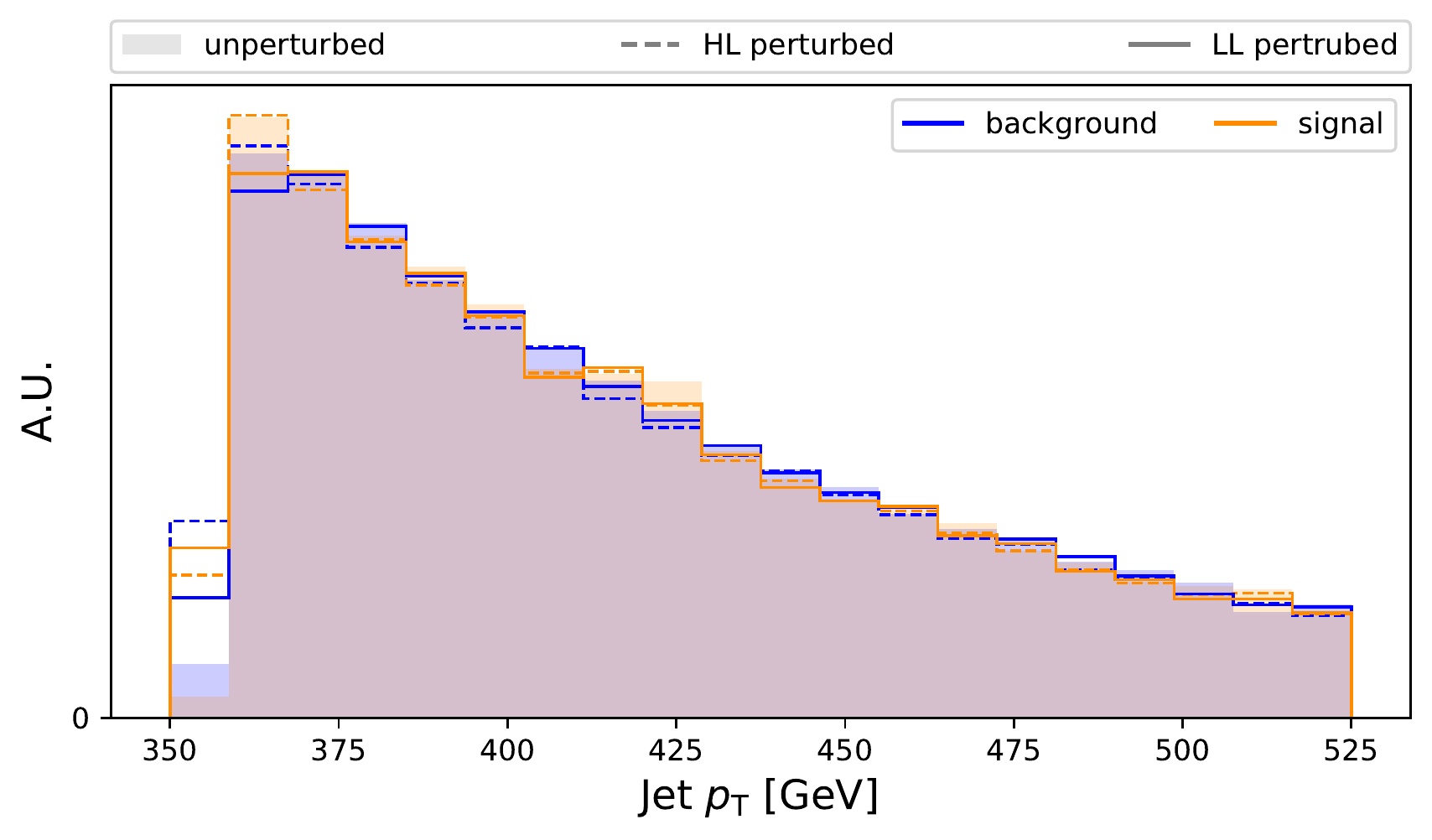} \\
\includegraphics[width=0.35\textwidth]{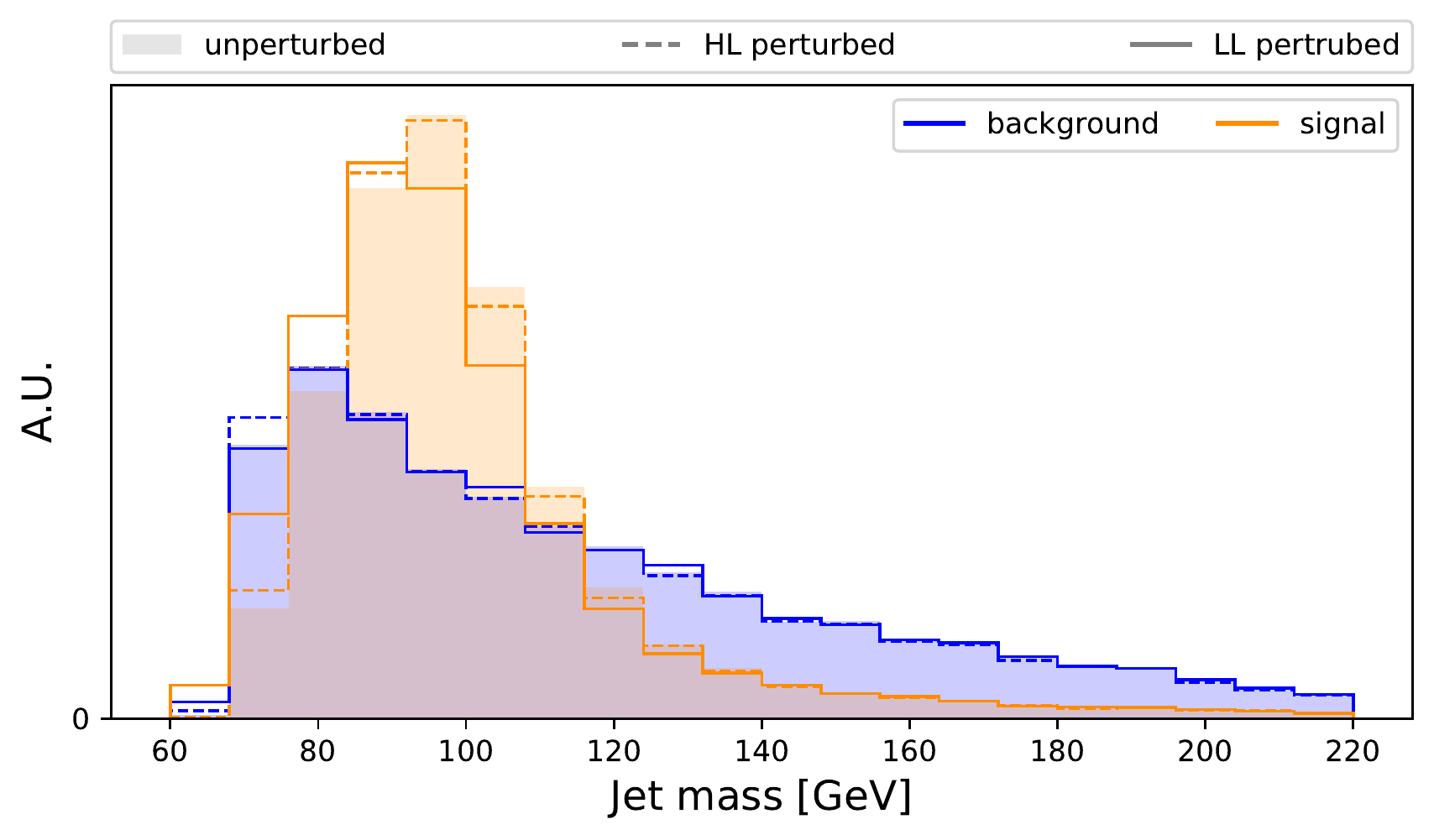} \\
\includegraphics[width=0.35\textwidth]{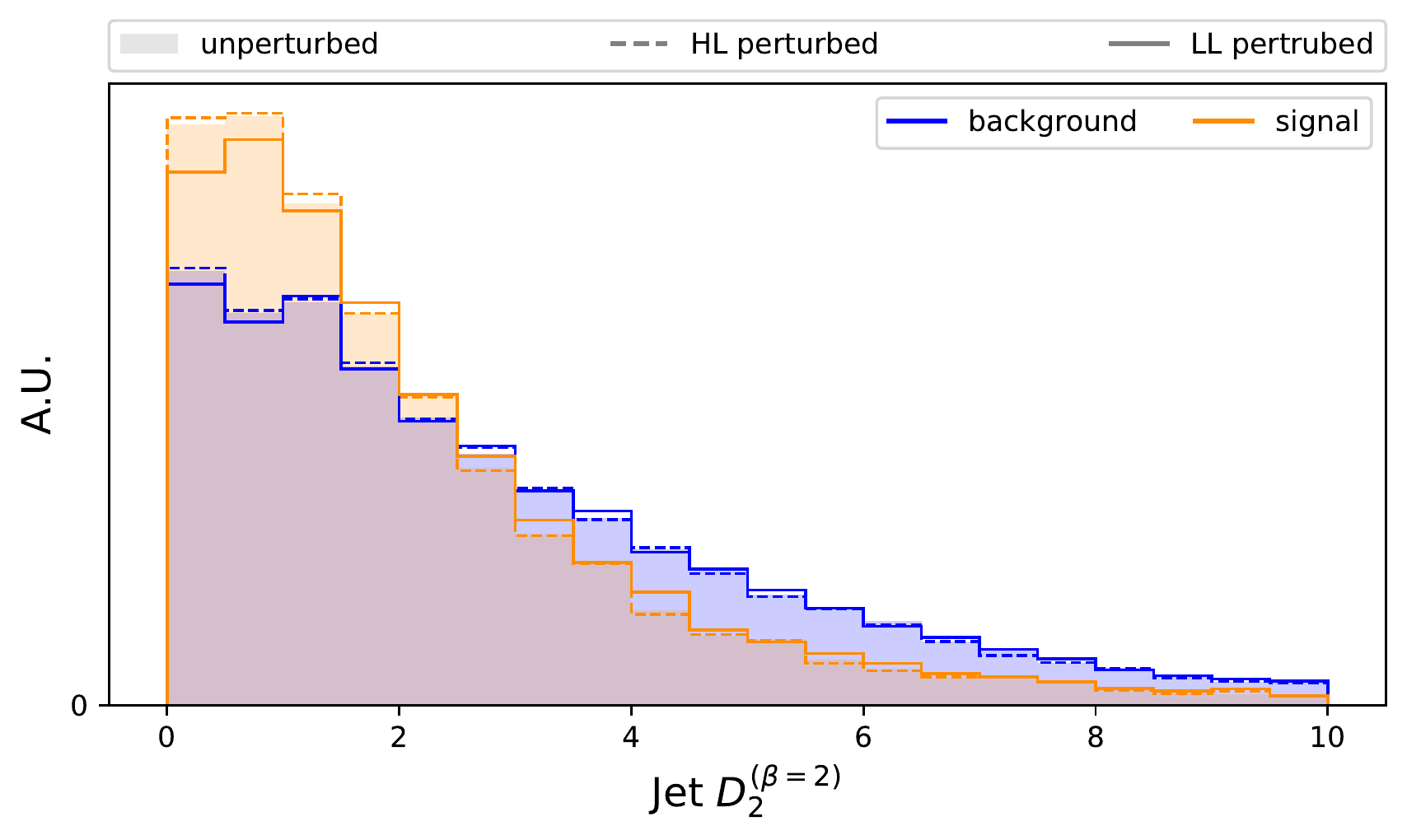}
\caption{
Comparison of the effect the adversarial network perturbations on the LL and HL classifier response, as well as various jet observables.
\label{fig:distributionsadversary}
}
\end{figure}

Representative HL features and the classifier distributions for the adversarial attack are presented in Fig.~\ref{fig:distributionsadversary}.
Even though both signal and background jets are subjected to the same adversary, the background distributions are nearly identical before and after the perturbation.
In contrast, the classifier response and mass distributions are noticeably distorted for the signal.
This allows the systematic mismodeling induced by the adversary to go undetected in typical experimental conditions, as shown in Fig.~\ref{fig:AdvValidation}.
The green line delineating the signal region corresponds to the maximum discovery significance expected based on the simulated signal and background models.
The shaded region, defined as the region in which expected signal efficiency exceeds 10\%, is taken to be blinded during experimental design and validation phase.
`Observations' are samples from the perturbed simulation and the `Expected' prediction is the unperturbed simulation.
The jet $\pt$ and mass distributions in the validation region agree well between the Observed and Expected values to within statistical uncertainty.
Despite this apparent agreement, due to the adversary's effect on jets in the signal region, the discovery sensitivity for a potential signal at the predicted optimal working point is reduced by about 25\% as shown in Fig.~\ref{fig:method2b}.

\section{Discussion}
\label{sec:discussion}

In the traditional HEP search paradigm, simulations are used to extrapolate predictions from a control region to a signal region.  Many sources of uncertainty on this extrapolation are well-constrained from auxiliary measurements on individual reconstructed objects, but others are mostly unvalidated from data.  For example, the modeling of strong-force processes related to hadron formation is a complex multiscale process for which it is customary to compare two different models (such as Pythia~\cite{Sjostrand:2006za,Sjostrand:2007gs} and Herwig~\cite{Bahr:2008pv,Bellm:2015jjp} or Sherpa~\cite{Gleisberg:2008ta,Bothmann:2019yzt}).
These algorithms model the same physical processes in different ways, with a mix of formal and phenomenological insight, and the difference between models is treated as a systematic uncertainty in statistical analyses.
Given the high dimensionality of collider events, it is unlikely that a single nuisance parameter encoding the distance between two arbitrary models represents a reasonable prior to cover the distribution of all possible systematic mismodelings.

The results of Sec.~\ref{sec:results} show that small perturbations in the high-dimensional phase space of collider events can significantly change the scientific conclusions from a given dataset.  If a complete uncertainty model in high-dimensions ensured that such perturbations were unphysical, then the adversarial results are irrelevant.  However, this is not the current situation described above.  Furthermore, networks trained on high-dimensional low-level inputs can outperform networks trained on high-level features precisely because they can take advantage of subtle correlations distributed across multiple dimensions.  The modeling of such correlations are particularly difficult to validate.  Significant physics input is required to build a full phase space uncertainty model. 

As is characteristic of problems involving nonconvex optimization, is difficult to show that a particular adversarial attack is maximal.
Therefore, the existence of an attack with a particular impact only provides a lower bound for the upper bound.
Nonetheless, if a certain network architecture is repeatedly shown to be particuarly difficult to attack, it may lend credibility to the currently-accepted treatment of simply ignoring certain high-dimensional systematic uncertainties.

In any case, for now the method described here offers the only rigorous means for quantifying how sensitive an analysis procedure is to high-dimensional mismodeling.
A given analysis procedure, including the control/signal region definition and any auxiliary features that will be validated in data, can be attacked to quantify the impact on the signal sensitivity.
While general methods from AI safety may also be useful for making classifiers robust to attacks, Sec.~\ref{sec:results} demonstrated that networks based on physically-motivated features can be less sensitive to adversarial perturbations, if only because it is feasible to ensure these observables are modeled reliably. %
However, it may be possible to design sensitive architectures that are able to leverage low-level information while remaining robust against adversarial attacks, for example by exploiting symmetries and other physically-motivated constraints.
Additionally, evidence suggests that purposefully undertraining HDLL networks may serve to reduce systematic exposure while meeting the performance of simpler HL network models.

Even though the adversarial methods presented in Sec.~\ref{sec:results} were able to make targeted attacks knowing the full form of the classifier, they are not the most general attack possible.  First of all, the perturbations were not allowed to split particles into multiple particles nor were they able to add new particles.  The modeling of such `soft' and `collinear' physics is particularly challenging and so such effects are an interesting class of perturbations for future studies.  Second, the true values of individual features or combinations of features are not observable - only distribution-level statistics can be validated.  In this work, per-constituent perturbations were constrained to ensure that observable features were approximately unperturbed on a jet-by-jet basis.  However, it is worth considering adversarial examples which have limited resemblance to any particular jet in the original dataset, while preserving properties of ensembles of events (such as the jet mass distribution).  One may be able to generalize the procedures described here using constraints on sufficiently large mini-batches or even on entire datasets.

\section{Conclusions}

The interest in deep learning methods for HEP has grown significantly since the first studies were published five years ago~\cite{Baldi:2014kfa}.  While these methods hold great promise to enhance experimental sensitivity to discover new fundamental properties of nature, conventional analysis techniques must be updated.  We have shown that neural networks using high-dimensional low-level features (and to a lesser extent, high-dimensional high-level features) are highly sensitive to mismodeled inputs.  Current uncertainty estimates may not be sufficient to address uncertainties involved when using high-dimensional features, and traditional validation methods may be ineffective in detecting such problems.  We have proposed adversarial approaches to evaluate and compare the sensitivity of deep learning-based analysis procedures.  While this is a crude bound, it may be used to demonstrate robustness against specific classes of uncertainty, or to diagnose situations where further studies are needed.  This work will hopefully begin a dialogue within the community about the robust application of deep learning to experimental measurements and searches.

\section*{ACKNOWLEDGMENTS}

We thank Paul Tipton for constructive feedback since the early stages of this project.
We are also grateful to Jesse Thaler and Daniel Whiteson for insightful comments and discussions.
This work is supported by the DOE under contracts DE-AC02-05CH11231 and DE-SC0017660.  BPN would like to thank NVIDIA for providing Volta GPUs used for some of the numerical examples and the Aspen Center for Physics, which is supported by National Science Foundation grant PHY-1607611.

\bibliography{myrefs}

\begin{thebibliography}{34}%
\makeatletter
\providecommand \@ifxundefined [1]{%
 \@ifx{#1\undefined}
}%
\providecommand \@ifnum [1]{%
 \ifnum #1\expandafter \@firstoftwo
 \else \expandafter \@secondoftwo
 \fi
}%
\providecommand \@ifx [1]{%
 \ifx #1\expandafter \@firstoftwo
 \else \expandafter \@secondoftwo
 \fi
}%
\providecommand \natexlab [1]{#1}%
\providecommand \enquote  [1]{``#1''}%
\providecommand \bibnamefont  [1]{#1}%
\providecommand \bibfnamefont [1]{#1}%
\providecommand \citenamefont [1]{#1}%
\providecommand \href@noop [0]{\@secondoftwo}%
\providecommand \href [0]{\begingroup \@sanitize@url \@href}%
\providecommand \@href[1]{\@@startlink{#1}\@@href}%
\providecommand \@@href[1]{\endgroup#1\@@endlink}%
\providecommand \@sanitize@url [0]{\catcode `\\12\catcode `\$12\catcode
  `\&12\catcode `\#12\catcode `\^12\catcode `\_12\catcode `\%12\relax}%
\providecommand \@@startlink[1]{}%
\providecommand \@@endlink[0]{}%
\providecommand \url  [0]{\begingroup\@sanitize@url \@url }%
\providecommand \@url [1]{\endgroup\@href {#1}{\urlprefix }}%
\providecommand \urlprefix  [0]{URL }%
\providecommand \Eprint [0]{\href }%
\providecommand \doibase [0]{http://dx.doi.org/}%
\providecommand \selectlanguage [0]{\@gobble}%
\providecommand \bibinfo  [0]{\@secondoftwo}%
\providecommand \bibfield  [0]{\@secondoftwo}%
\providecommand \translation [1]{[#1]}%
\providecommand \BibitemOpen [0]{}%
\providecommand \bibitemStop [0]{}%
\providecommand \bibitemNoStop [0]{.\EOS\space}%
\providecommand \EOS [0]{\spacefactor3000\relax}%
\providecommand \BibitemShut  [1]{\csname bibitem#1\endcsname}%
\let\auto@bib@innerbib\@empty
%</preamble>
\bibitem [{\citenamefont {Larkoski}\ \emph {et~al.}(2017)\citenamefont
  {Larkoski}, \citenamefont {Moult},\ and\ \citenamefont
  {Nachman}}]{Larkoski:2017jix}%
  \BibitemOpen
  \bibfield  {author} {\bibinfo {author} {\bibfnamefont {A.~J.}\ \bibnamefont
  {Larkoski}}, \bibinfo {author} {\bibfnamefont {I.}~\bibnamefont {Moult}}, \
  and\ \bibinfo {author} {\bibfnamefont {B.}~\bibnamefont {Nachman}},\
  }\href@noop {} {\  (\bibinfo {year} {2017})},\ \Eprint
  {http://arxiv.org/abs/1709.04464} {arXiv:1709.04464 [hep-ph]} \BibitemShut
  {NoStop}%
%%CITATION = ARXIV:1709.04464;%%
\bibitem [{\citenamefont {Radovic}\ \emph {et~al.}(2018)\citenamefont
  {Radovic}, \citenamefont {Williams}, \citenamefont {Rousseau}, \citenamefont
  {Kagan}, \citenamefont {Bonacorsi}, \citenamefont {Himmel}, \citenamefont
  {Aurisano}, \citenamefont {Terao},\ and\ \citenamefont
  {Wongjirad}}]{Radovic:2018dip}%
  \BibitemOpen
  \bibfield  {author} {\bibinfo {author} {\bibfnamefont {A.}~\bibnamefont
  {Radovic}}, \bibinfo {author} {\bibfnamefont {M.}~\bibnamefont {Williams}},
  \bibinfo {author} {\bibfnamefont {D.}~\bibnamefont {Rousseau}}, \bibinfo
  {author} {\bibfnamefont {M.}~\bibnamefont {Kagan}}, \bibinfo {author}
  {\bibfnamefont {D.}~\bibnamefont {Bonacorsi}}, \bibinfo {author}
  {\bibfnamefont {A.}~\bibnamefont {Himmel}}, \bibinfo {author} {\bibfnamefont
  {A.}~\bibnamefont {Aurisano}}, \bibinfo {author} {\bibfnamefont
  {K.}~\bibnamefont {Terao}}, \ and\ \bibinfo {author} {\bibfnamefont
  {T.}~\bibnamefont {Wongjirad}},\ }\href {\doibase 10.1038/s41586-018-0361-2}
  {\bibfield  {journal} {\bibinfo  {journal} {Nature}\ }\textbf {\bibinfo
  {volume} {560}},\ \bibinfo {pages} {41} (\bibinfo {year} {2018})}\BibitemShut
  {NoStop}%
%%CITATION = NATUA,560,41;%%
\bibitem [{\citenamefont {Guest}\ \emph {et~al.}(2018)\citenamefont {Guest},
  \citenamefont {Cranmer},\ and\ \citenamefont {Whiteson}}]{Guest:2018yhq}%
  \BibitemOpen
  \bibfield  {author} {\bibinfo {author} {\bibfnamefont {D.}~\bibnamefont
  {Guest}}, \bibinfo {author} {\bibfnamefont {K.}~\bibnamefont {Cranmer}}, \
  and\ \bibinfo {author} {\bibfnamefont {D.}~\bibnamefont {Whiteson}},\ }\href
  {\doibase 10.1146/annurev-nucl-101917-021019} {\bibfield  {journal} {\bibinfo
   {journal} {Ann. Rev. Nucl. Part. Sci.}\ }\textbf {\bibinfo {volume} {68}},\
  \bibinfo {pages} {161} (\bibinfo {year} {2018})},\ \Eprint
  {http://arxiv.org/abs/1806.11484} {arXiv:1806.11484 [hep-ex]} \BibitemShut
  {NoStop}%
%%CITATION = ARXIV:1806.11484;%%
\bibitem [{\citenamefont {Baldi}\ \emph {et~al.}(2014)\citenamefont {Baldi},
  \citenamefont {Sadowski},\ and\ \citenamefont {Whiteson}}]{Baldi:2014kfa}%
  \BibitemOpen
  \bibfield  {author} {\bibinfo {author} {\bibfnamefont {P.}~\bibnamefont
  {Baldi}}, \bibinfo {author} {\bibfnamefont {P.}~\bibnamefont {Sadowski}}, \
  and\ \bibinfo {author} {\bibfnamefont {D.}~\bibnamefont {Whiteson}},\ }\href
  {\doibase 10.1038/ncomms5308} {\bibfield  {journal} {\bibinfo  {journal}
  {Nature Commun.}\ }\textbf {\bibinfo {volume} {5}},\ \bibinfo {pages} {4308}
  (\bibinfo {year} {2014})},\ \Eprint {http://arxiv.org/abs/1402.4735}
  {arXiv:1402.4735 [hep-ph]} \BibitemShut {NoStop}%
%%CITATION = ARXIV:1402.4735;%%
\bibitem [{\citenamefont {de~Oliveira}\ \emph {et~al.}(2016)\citenamefont
  {de~Oliveira}, \citenamefont {Kagan}, \citenamefont {Mackey}, \citenamefont
  {Nachman},\ and\ \citenamefont {Schwartzman}}]{deOliveira:2015xxd}%
  \BibitemOpen
  \bibfield  {author} {\bibinfo {author} {\bibfnamefont {L.}~\bibnamefont
  {de~Oliveira}}, \bibinfo {author} {\bibfnamefont {M.}~\bibnamefont {Kagan}},
  \bibinfo {author} {\bibfnamefont {L.}~\bibnamefont {Mackey}}, \bibinfo
  {author} {\bibfnamefont {B.}~\bibnamefont {Nachman}}, \ and\ \bibinfo
  {author} {\bibfnamefont {A.}~\bibnamefont {Schwartzman}},\ }\href {\doibase
  10.1007/JHEP07(2016)069} {\bibfield  {journal} {\bibinfo  {journal} {JHEP}\
  }\textbf {\bibinfo {volume} {07}},\ \bibinfo {pages} {069} (\bibinfo {year}
  {2016})},\ \Eprint {http://arxiv.org/abs/1511.05190} {arXiv:1511.05190
  [hep-ph]} \BibitemShut {NoStop}%
%%CITATION = ARXIV:1511.05190;%%
\bibitem [{\citenamefont {Baldi}\ \emph {et~al.}(2016)\citenamefont {Baldi},
  \citenamefont {Bauer}, \citenamefont {Eng}, \citenamefont {Sadowski},\ and\
  \citenamefont {Whiteson}}]{Baldi:2016fql}%
  \BibitemOpen
  \bibfield  {author} {\bibinfo {author} {\bibfnamefont {P.}~\bibnamefont
  {Baldi}}, \bibinfo {author} {\bibfnamefont {K.}~\bibnamefont {Bauer}},
  \bibinfo {author} {\bibfnamefont {C.}~\bibnamefont {Eng}}, \bibinfo {author}
  {\bibfnamefont {P.}~\bibnamefont {Sadowski}}, \ and\ \bibinfo {author}
  {\bibfnamefont {D.}~\bibnamefont {Whiteson}},\ }\href {\doibase
  10.1103/PhysRevD.93.094034} {\bibfield  {journal} {\bibinfo  {journal} {Phys.
  Rev.}\ }\textbf {\bibinfo {volume} {D93}},\ \bibinfo {pages} {094034}
  (\bibinfo {year} {2016})},\ \Eprint {http://arxiv.org/abs/1603.09349}
  {arXiv:1603.09349 [hep-ex]} \BibitemShut {NoStop}%
%%CITATION = ARXIV:1603.09349;%%
\bibitem [{\citenamefont {Guest}\ \emph {et~al.}(2016)\citenamefont {Guest},
  \citenamefont {Collado}, \citenamefont {Baldi}, \citenamefont {Hsu},
  \citenamefont {Urban},\ and\ \citenamefont {Whiteson}}]{Guest:2016iqz}%
  \BibitemOpen
  \bibfield  {author} {\bibinfo {author} {\bibfnamefont {D.}~\bibnamefont
  {Guest}}, \bibinfo {author} {\bibfnamefont {J.}~\bibnamefont {Collado}},
  \bibinfo {author} {\bibfnamefont {P.}~\bibnamefont {Baldi}}, \bibinfo
  {author} {\bibfnamefont {S.-C.}\ \bibnamefont {Hsu}}, \bibinfo {author}
  {\bibfnamefont {G.}~\bibnamefont {Urban}}, \ and\ \bibinfo {author}
  {\bibfnamefont {D.}~\bibnamefont {Whiteson}},\ }\href {\doibase
  10.1103/PhysRevD.94.112002} {\bibfield  {journal} {\bibinfo  {journal} {Phys.
  Rev.}\ }\textbf {\bibinfo {volume} {D94}},\ \bibinfo {pages} {112002}
  (\bibinfo {year} {2016})},\ \Eprint {http://arxiv.org/abs/1607.08633}
  {arXiv:1607.08633 [hep-ex]} \BibitemShut {NoStop}%
%%CITATION = ARXIV:1607.08633;%%
\bibitem [{\citenamefont {{ATLAS
  Collaboration}}(2019{\natexlab{a}})}]{Aad:2019yxi}%
  \BibitemOpen
  \bibfield  {author} {\bibinfo {author} {\bibnamefont {{ATLAS
  Collaboration}}},\ }\href@noop {} {\  (\bibinfo {year}
  {2019}{\natexlab{a}})},\ \Eprint {http://arxiv.org/abs/1908.06765}
  {arXiv:1908.06765 [hep-ex]} \BibitemShut {NoStop}%
%%CITATION = ARXIV:1908.06765;%%
\bibitem [{\citenamefont {{ATLAS
  Collaboration}}(2019{\natexlab{b}})}]{ATLAS-CONF-2019-017}%
  \BibitemOpen
  \bibfield  {author} {\bibinfo {author} {\bibnamefont {{ATLAS
  Collaboration}}},\ }\href {http://cds.cern.ch/record/2676594} {\bibfield
  {journal} {\bibinfo  {journal} {ATLAS-CONF-2019-017}\ } (\bibinfo {year}
  {2019}{\natexlab{b}})}\BibitemShut {NoStop}%
\bibitem [{\citenamefont {{CMS Collaboration}}(2019)}]{CMS-PAS-SUS-19-009}%
  \BibitemOpen
  \bibfield  {author} {\bibinfo {author} {\bibnamefont {{CMS Collaboration}}},\
  }\href {http://cds.cern.ch/record/2682157} {\bibfield  {journal} {\bibinfo
  {journal} {CMS-PAS-SUS-19-009}\ } (\bibinfo {year} {2019})}\BibitemShut
  {NoStop}%
\bibitem [{\citenamefont {{CMS Collaboration}}(2017)}]{CMS-DP-2017-005}%
  \BibitemOpen
  \bibfield  {author} {\bibinfo {author} {\bibnamefont {{CMS Collaboration}}},\
  }\href {https://cds.cern.ch/record/2255736} {\bibfield  {journal} {\bibinfo
  {journal} {CMS-DP-2017-005}\ } (\bibinfo {year} {2017})}\BibitemShut
  {NoStop}%
\bibitem [{\citenamefont {{ATLAS
  Collaboration}}(2017)}]{ATL-PHYS-PUB-2017-003}%
  \BibitemOpen
  \bibfield  {author} {\bibinfo {author} {\bibnamefont {{ATLAS
  Collaboration}}},\ }\href {http://cds.cern.ch/record/2255226} {\bibfield
  {journal} {\bibinfo  {journal} {ATL-PHYS-PUB-2017-003}\ } (\bibinfo {year}
  {2017})}\BibitemShut {NoStop}%
\bibitem [{\citenamefont {Andrews}\ \emph {et~al.}(2018)\citenamefont
  {Andrews}, \citenamefont {Paulini}, \citenamefont {Gleyzer},\ and\
  \citenamefont {Poczos}}]{Andrews:2018nwy}%
  \BibitemOpen
  \bibfield  {author} {\bibinfo {author} {\bibfnamefont {M.}~\bibnamefont
  {Andrews}}, \bibinfo {author} {\bibfnamefont {M.}~\bibnamefont {Paulini}},
  \bibinfo {author} {\bibfnamefont {S.}~\bibnamefont {Gleyzer}}, \ and\
  \bibinfo {author} {\bibfnamefont {B.}~\bibnamefont {Poczos}},\ }\href@noop {}
  {\  (\bibinfo {year} {2018})},\ \Eprint {http://arxiv.org/abs/1807.11916}
  {arXiv:1807.11916 [hep-ex]} \BibitemShut {NoStop}%
%%CITATION = ARXIV:1807.11916;%%
\bibitem [{\citenamefont {Andrews}\ \emph {et~al.}(2019)\citenamefont
  {Andrews}, \citenamefont {Alison}, \citenamefont {An}, \citenamefont
  {Bryant}, \citenamefont {Burkle}, \citenamefont {Gleyzer}, \citenamefont
  {Narain}, \citenamefont {Paulini}, \citenamefont {Poczos},\ and\
  \citenamefont {Usai}}]{Andrews:2019faz}%
  \BibitemOpen
  \bibfield  {author} {\bibinfo {author} {\bibfnamefont {M.}~\bibnamefont
  {Andrews}}, \bibinfo {author} {\bibfnamefont {J.}~\bibnamefont {Alison}},
  \bibinfo {author} {\bibfnamefont {S.}~\bibnamefont {An}}, \bibinfo {author}
  {\bibfnamefont {P.}~\bibnamefont {Bryant}}, \bibinfo {author} {\bibfnamefont
  {B.}~\bibnamefont {Burkle}}, \bibinfo {author} {\bibfnamefont
  {S.}~\bibnamefont {Gleyzer}}, \bibinfo {author} {\bibfnamefont
  {M.}~\bibnamefont {Narain}}, \bibinfo {author} {\bibfnamefont
  {M.}~\bibnamefont {Paulini}}, \bibinfo {author} {\bibfnamefont
  {B.}~\bibnamefont {Poczos}}, \ and\ \bibinfo {author} {\bibfnamefont
  {E.}~\bibnamefont {Usai}},\ }\href@noop {} {\  (\bibinfo {year} {2019})},\
  \Eprint {http://arxiv.org/abs/1902.08276} {arXiv:1902.08276 [hep-ex]}
  \BibitemShut {NoStop}%
%%CITATION = ARXIV:1902.08276;%%
\bibitem [{\citenamefont {Nachman}(2019)}]{Nachman:2019dol}%
  \BibitemOpen
  \bibfield  {author} {\bibinfo {author} {\bibfnamefont {B.}~\bibnamefont
  {Nachman}},\ }\href@noop {} {\  (\bibinfo {year} {2019})},\ \Eprint
  {http://arxiv.org/abs/1909.03081} {arXiv:1909.03081 [hep-ph]} \BibitemShut
  {NoStop}%
%%CITATION = ARXIV:1909.03081;%%
\bibitem [{\citenamefont {Szegedy}\ \emph {et~al.}(2014)\citenamefont
  {Szegedy}, \citenamefont {Inc}, \citenamefont {Zaremba}, \citenamefont
  {Sutskever}, \citenamefont {Inc}, \citenamefont {Bruna}, \citenamefont
  {Erhan}, \citenamefont {Inc}, \citenamefont {Goodfellow},\ and\ \citenamefont
  {Fergus}}]{Szegedy14intriguingproperties}%
  \BibitemOpen
  \bibfield  {author} {\bibinfo {author} {\bibfnamefont {C.}~\bibnamefont
  {Szegedy}}, \bibinfo {author} {\bibfnamefont {G.}~\bibnamefont {Inc}},
  \bibinfo {author} {\bibfnamefont {W.}~\bibnamefont {Zaremba}}, \bibinfo
  {author} {\bibfnamefont {I.}~\bibnamefont {Sutskever}}, \bibinfo {author}
  {\bibfnamefont {G.}~\bibnamefont {Inc}}, \bibinfo {author} {\bibfnamefont
  {J.}~\bibnamefont {Bruna}}, \bibinfo {author} {\bibfnamefont
  {D.}~\bibnamefont {Erhan}}, \bibinfo {author} {\bibfnamefont
  {G.}~\bibnamefont {Inc}}, \bibinfo {author} {\bibfnamefont {I.}~\bibnamefont
  {Goodfellow}}, \ and\ \bibinfo {author} {\bibfnamefont {R.}~\bibnamefont
  {Fergus}},\ }in\ \href@noop {} {\emph {\bibinfo {booktitle} {In ICLR}}}\
  (\bibinfo {year} {2014})\BibitemShut {NoStop}%
\bibitem [{\citenamefont {Goodfellow}\ \emph {et~al.}(2015)\citenamefont
  {Goodfellow}, \citenamefont {Shlens},\ and\ \citenamefont
  {Szegedy}}]{DBLP:journals/corr/GoodfellowSS14}%
  \BibitemOpen
  \bibfield  {author} {\bibinfo {author} {\bibfnamefont {I.~J.}\ \bibnamefont
  {Goodfellow}}, \bibinfo {author} {\bibfnamefont {J.}~\bibnamefont {Shlens}},
  \ and\ \bibinfo {author} {\bibfnamefont {C.}~\bibnamefont {Szegedy}},\ }in\
  \href {http://arxiv.org/abs/1412.6572} {\emph {\bibinfo {booktitle} {3rd
  International Conference on Learning Representations, {ICLR} 2015, San Diego,
  CA, USA, May 7-9, 2015, Conference Track Proceedings}}}\ (\bibinfo {year}
  {2015})\BibitemShut {NoStop}%
\bibitem [{\citenamefont {THOMSON}(1874)}]{TheLordKelvin}%
  \BibitemOpen
  \bibfield  {author} {\bibinfo {author} {\bibfnamefont {W.}~\bibnamefont
  {THOMSON}},\ }\href {\doibase 10.1038/009441c0} {\bibfield  {journal}
  {\bibinfo  {journal} {Nature}\ }\textbf {\bibinfo {volume} {9}},\ \bibinfo
  {pages} {441} (\bibinfo {year} {1874})}\BibitemShut {NoStop}%
\bibitem [{\citenamefont {Cacciari}\ \emph {et~al.}(2008)\citenamefont
  {Cacciari}, \citenamefont {Salam},\ and\ \citenamefont
  {Soyez}}]{Cacciari:2008gp}%
  \BibitemOpen
  \bibfield  {author} {\bibinfo {author} {\bibfnamefont {M.}~\bibnamefont
  {Cacciari}}, \bibinfo {author} {\bibfnamefont {G.~P.}\ \bibnamefont {Salam}},
  \ and\ \bibinfo {author} {\bibfnamefont {G.}~\bibnamefont {Soyez}},\ }\href
  {\doibase 10.1088/1126-6708/2008/04/063} {\bibfield  {journal} {\bibinfo
  {journal} {JHEP}\ }\textbf {\bibinfo {volume} {04}},\ \bibinfo {pages} {063}
  (\bibinfo {year} {2008})},\ \Eprint {http://arxiv.org/abs/0802.1189}
  {arXiv:0802.1189 [hep-ph]} \BibitemShut {NoStop}%
%%CITATION = ARXIV:0802.1189;%%
\bibitem [{\citenamefont {Alwall}\ \emph {et~al.}(2014)\citenamefont {Alwall},
  \citenamefont {Frederix}, \citenamefont {Frixione}, \citenamefont {Hirschi},
  \citenamefont {Maltoni}, \citenamefont {Mattelaer}, \citenamefont {Shao},
  \citenamefont {Stelzer}, \citenamefont {Torrielli},\ and\ \citenamefont
  {Zaro}}]{Alwall:2014hca}%
  \BibitemOpen
  \bibfield  {author} {\bibinfo {author} {\bibfnamefont {J.}~\bibnamefont
  {Alwall}}, \bibinfo {author} {\bibfnamefont {R.}~\bibnamefont {Frederix}},
  \bibinfo {author} {\bibfnamefont {S.}~\bibnamefont {Frixione}}, \bibinfo
  {author} {\bibfnamefont {V.}~\bibnamefont {Hirschi}}, \bibinfo {author}
  {\bibfnamefont {F.}~\bibnamefont {Maltoni}}, \bibinfo {author} {\bibfnamefont
  {O.}~\bibnamefont {Mattelaer}}, \bibinfo {author} {\bibfnamefont {H.~S.}\
  \bibnamefont {Shao}}, \bibinfo {author} {\bibfnamefont {T.}~\bibnamefont
  {Stelzer}}, \bibinfo {author} {\bibfnamefont {P.}~\bibnamefont {Torrielli}},
  \ and\ \bibinfo {author} {\bibfnamefont {M.}~\bibnamefont {Zaro}},\ }\href
  {\doibase 10.1007/JHEP07(2014)079} {\bibfield  {journal} {\bibinfo  {journal}
  {JHEP}\ }\textbf {\bibinfo {volume} {07}},\ \bibinfo {pages} {079} (\bibinfo
  {year} {2014})},\ \Eprint {http://arxiv.org/abs/1405.0301} {arXiv:1405.0301
  [hep-ph]} \BibitemShut {NoStop}%
%%CITATION = ARXIV:1405.0301;%%
\bibitem [{\citenamefont {Sjostrand}\ \emph {et~al.}(2006)\citenamefont
  {Sjostrand}, \citenamefont {Mrenna},\ and\ \citenamefont
  {Skands}}]{Sjostrand:2006za}%
  \BibitemOpen
  \bibfield  {author} {\bibinfo {author} {\bibfnamefont {T.}~\bibnamefont
  {Sjostrand}}, \bibinfo {author} {\bibfnamefont {S.}~\bibnamefont {Mrenna}}, \
  and\ \bibinfo {author} {\bibfnamefont {P.~Z.}\ \bibnamefont {Skands}},\
  }\href {\doibase 10.1088/1126-6708/2006/05/026} {\bibfield  {journal}
  {\bibinfo  {journal} {JHEP}\ }\textbf {\bibinfo {volume} {05}},\ \bibinfo
  {pages} {026} (\bibinfo {year} {2006})},\ \Eprint
  {http://arxiv.org/abs/hep-ph/0603175} {arXiv:hep-ph/0603175 [hep-ph]}
  \BibitemShut {NoStop}%
%%CITATION = HEP-PH/0603175;%%
\bibitem [{\citenamefont {Sjostrand}\ \emph {et~al.}(2008)\citenamefont
  {Sjostrand}, \citenamefont {Mrenna},\ and\ \citenamefont
  {Skands}}]{Sjostrand:2007gs}%
  \BibitemOpen
  \bibfield  {author} {\bibinfo {author} {\bibfnamefont {T.}~\bibnamefont
  {Sjostrand}}, \bibinfo {author} {\bibfnamefont {S.}~\bibnamefont {Mrenna}}, \
  and\ \bibinfo {author} {\bibfnamefont {P.~Z.}\ \bibnamefont {Skands}},\
  }\href {\doibase 10.1016/j.cpc.2008.01.036} {\bibfield  {journal} {\bibinfo
  {journal} {Comput. Phys. Commun.}\ }\textbf {\bibinfo {volume} {178}},\
  \bibinfo {pages} {852} (\bibinfo {year} {2008})},\ \Eprint
  {http://arxiv.org/abs/0710.3820} {arXiv:0710.3820 [hep-ph]} \BibitemShut
  {NoStop}%
%%CITATION = ARXIV:0710.3820;%%
\bibitem [{\citenamefont {de~Favereau}\ \emph {et~al.}(2014)\citenamefont
  {de~Favereau}, \citenamefont {Delaere}, \citenamefont {Demin}, \citenamefont
  {Giammanco}, \citenamefont {Lemaitre}, \citenamefont {Mertens},\ and\
  \citenamefont {Selvaggi}}]{deFavereau:2013fsa}%
  \BibitemOpen
  \bibfield  {author} {\bibinfo {author} {\bibfnamefont {J.}~\bibnamefont
  {de~Favereau}}, \bibinfo {author} {\bibfnamefont {C.}~\bibnamefont
  {Delaere}}, \bibinfo {author} {\bibfnamefont {P.}~\bibnamefont {Demin}},
  \bibinfo {author} {\bibfnamefont {A.}~\bibnamefont {Giammanco}}, \bibinfo
  {author} {\bibfnamefont {V.}~\bibnamefont {Lemaitre}}, \bibinfo {author}
  {\bibfnamefont {A.}~\bibnamefont {Mertens}}, \ and\ \bibinfo {author}
  {\bibfnamefont {M.}~\bibnamefont {Selvaggi}} (\bibinfo {collaboration}
  {DELPHES 3}),\ }\href {\doibase 10.1007/JHEP02(2014)057} {\bibfield
  {journal} {\bibinfo  {journal} {JHEP}\ }\textbf {\bibinfo {volume} {02}},\
  \bibinfo {pages} {057} (\bibinfo {year} {2014})},\ \Eprint
  {http://arxiv.org/abs/1307.6346} {arXiv:1307.6346 [hep-ex]} \BibitemShut
  {NoStop}%
%%CITATION = ARXIV:1307.6346;%%
\bibitem [{\citenamefont {Larkoski}\ \emph {et~al.}(2014)\citenamefont
  {Larkoski}, \citenamefont {Moult},\ and\ \citenamefont
  {Neill}}]{Larkoski:2014gra}%
  \BibitemOpen
  \bibfield  {author} {\bibinfo {author} {\bibfnamefont {A.~J.}\ \bibnamefont
  {Larkoski}}, \bibinfo {author} {\bibfnamefont {I.}~\bibnamefont {Moult}}, \
  and\ \bibinfo {author} {\bibfnamefont {D.}~\bibnamefont {Neill}},\ }\href
  {\doibase 10.1007/JHEP12(2014)009} {\bibfield  {journal} {\bibinfo  {journal}
  {JHEP}\ }\textbf {\bibinfo {volume} {12}},\ \bibinfo {pages} {009} (\bibinfo
  {year} {2014})},\ \Eprint {http://arxiv.org/abs/1409.6298} {arXiv:1409.6298
  [hep-ph]} \BibitemShut {NoStop}%
%%CITATION = ARXIV:1409.6298;%%
\bibitem [{\citenamefont {Komiske}\ \emph {et~al.}(2019)\citenamefont
  {Komiske}, \citenamefont {Metodiev},\ and\ \citenamefont
  {Thaler}}]{Komiske:2018cqr}%
  \BibitemOpen
  \bibfield  {author} {\bibinfo {author} {\bibfnamefont {P.~T.}\ \bibnamefont
  {Komiske}}, \bibinfo {author} {\bibfnamefont {E.~M.}\ \bibnamefont
  {Metodiev}}, \ and\ \bibinfo {author} {\bibfnamefont {J.}~\bibnamefont
  {Thaler}},\ }\href {\doibase 10.1007/JHEP01(2019)121} {\bibfield  {journal}
  {\bibinfo  {journal} {JHEP}\ }\textbf {\bibinfo {volume} {01}},\ \bibinfo
  {pages} {121} (\bibinfo {year} {2019})},\ \Eprint
  {http://arxiv.org/abs/1810.05165} {arXiv:1810.05165 [hep-ph]} \BibitemShut
  {NoStop}%
%%CITATION = ARXIV:1810.05165;%%
\bibitem [{\citenamefont {Butter}\ \emph {et~al.}(2019)\citenamefont {Butter}
  \emph {et~al.}}]{Kasieczka:2019dbj}%
  \BibitemOpen
  \bibfield  {author} {\bibinfo {author} {\bibfnamefont {A.}~\bibnamefont
  {Butter}} \emph {et~al.},\ }\href {\doibase 10.21468/SciPostPhys.7.1.014}
  {\bibfield  {journal} {\bibinfo  {journal} {SciPost Phys.}\ }\textbf
  {\bibinfo {volume} {7}},\ \bibinfo {pages} {014} (\bibinfo {year} {2019})},\
  \Eprint {http://arxiv.org/abs/1902.09914} {arXiv:1902.09914 [hep-ph]}
  \BibitemShut {NoStop}%
%%CITATION = ARXIV:1902.09914;%%
\bibitem [{\citenamefont {Chollet}(2017)}]{keras}%
  \BibitemOpen
  \bibfield  {author} {\bibinfo {author} {\bibfnamefont {F.}~\bibnamefont
  {Chollet}},\ }\href@noop {} {\enquote {\bibinfo {title} {Keras},}\ }\bibinfo
  {howpublished} {\url{https://github.com/fchollet/keras}} (\bibinfo {year}
  {2017})\BibitemShut {NoStop}%
\bibitem [{\citenamefont {Abadi}\ \emph {et~al.}(2016)\citenamefont {Abadi},
  \citenamefont {Barham}, \citenamefont {Chen}, \citenamefont {Chen},
  \citenamefont {Davis}, \citenamefont {Dean}, \citenamefont {Devin},
  \citenamefont {Ghemawat}, \citenamefont {Irving}, \citenamefont {Isard} \emph
  {et~al.}}]{tensorflow}%
  \BibitemOpen
  \bibfield  {author} {\bibinfo {author} {\bibfnamefont {M.}~\bibnamefont
  {Abadi}}, \bibinfo {author} {\bibfnamefont {P.}~\bibnamefont {Barham}},
  \bibinfo {author} {\bibfnamefont {J.}~\bibnamefont {Chen}}, \bibinfo {author}
  {\bibfnamefont {Z.}~\bibnamefont {Chen}}, \bibinfo {author} {\bibfnamefont
  {A.}~\bibnamefont {Davis}}, \bibinfo {author} {\bibfnamefont
  {J.}~\bibnamefont {Dean}}, \bibinfo {author} {\bibfnamefont {M.}~\bibnamefont
  {Devin}}, \bibinfo {author} {\bibfnamefont {S.}~\bibnamefont {Ghemawat}},
  \bibinfo {author} {\bibfnamefont {G.}~\bibnamefont {Irving}}, \bibinfo
  {author} {\bibfnamefont {M.}~\bibnamefont {Isard}},  \emph {et~al.},\ }in\
  \href@noop {} {\emph {\bibinfo {booktitle} {OSDI}}},\ Vol.~\bibinfo {volume}
  {16}\ (\bibinfo {year} {2016})\ pp.\ \bibinfo {pages} {265--283}\BibitemShut
  {NoStop}%
\bibitem [{\citenamefont {Kingma}\ and\ \citenamefont {Ba}(2014)}]{adam}%
  \BibitemOpen
  \bibfield  {author} {\bibinfo {author} {\bibfnamefont {D.}~\bibnamefont
  {Kingma}}\ and\ \bibinfo {author} {\bibfnamefont {J.}~\bibnamefont {Ba}},\
  }\href@noop {} {\  (\bibinfo {year} {2014})},\ \Eprint
  {http://arxiv.org/abs/1412.6980} {arXiv:1412.6980 [cs]} \BibitemShut
  {NoStop}%
\bibitem [{\citenamefont {Cowan}\ \emph {et~al.}(2011)\citenamefont {Cowan},
  \citenamefont {Cranmer}, \citenamefont {Gross},\ and\ \citenamefont
  {Vitells}}]{Cowan:2010js}%
  \BibitemOpen
  \bibfield  {author} {\bibinfo {author} {\bibfnamefont {G.}~\bibnamefont
  {Cowan}}, \bibinfo {author} {\bibfnamefont {K.}~\bibnamefont {Cranmer}},
  \bibinfo {author} {\bibfnamefont {E.}~\bibnamefont {Gross}}, \ and\ \bibinfo
  {author} {\bibfnamefont {O.}~\bibnamefont {Vitells}},\ }\href {\doibase
  10.1140/epjc/s10052-011-1554-0, 10.1140/epjc/s10052-013-2501-z} {\bibfield
  {journal} {\bibinfo  {journal} {Eur. Phys. J.}\ }\textbf {\bibinfo {volume}
  {C71}},\ \bibinfo {pages} {1554} (\bibinfo {year} {2011})},\ \bibinfo {note}
  {[Erratum: Eur. Phys. J.C73,2501(2013)]},\ \Eprint
  {http://arxiv.org/abs/1007.1727} {arXiv:1007.1727 [physics.data-an]}
  \BibitemShut {NoStop}%
%%CITATION = ARXIV:1007.1727;%%
\bibitem [{\citenamefont {Bahr}\ \emph {et~al.}(2008)\citenamefont {Bahr} \emph
  {et~al.}}]{Bahr:2008pv}%
  \BibitemOpen
  \bibfield  {author} {\bibinfo {author} {\bibfnamefont {M.}~\bibnamefont
  {Bahr}} \emph {et~al.},\ }\href {\doibase 10.1140/epjc/s10052-008-0798-9}
  {\bibfield  {journal} {\bibinfo  {journal} {Eur. Phys. J.}\ }\textbf
  {\bibinfo {volume} {C58}},\ \bibinfo {pages} {639} (\bibinfo {year}
  {2008})},\ \Eprint {http://arxiv.org/abs/0803.0883} {arXiv:0803.0883
  [hep-ph]} \BibitemShut {NoStop}%
%%CITATION = ARXIV:0803.0883;%%
\bibitem [{\citenamefont {Bellm}\ \emph {et~al.}(2016)\citenamefont {Bellm}
  \emph {et~al.}}]{Bellm:2015jjp}%
  \BibitemOpen
  \bibfield  {author} {\bibinfo {author} {\bibfnamefont {J.}~\bibnamefont
  {Bellm}} \emph {et~al.},\ }\href {\doibase 10.1140/epjc/s10052-016-4018-8}
  {\bibfield  {journal} {\bibinfo  {journal} {Eur. Phys. J.}\ }\textbf
  {\bibinfo {volume} {C76}},\ \bibinfo {pages} {196} (\bibinfo {year}
  {2016})},\ \Eprint {http://arxiv.org/abs/1512.01178} {arXiv:1512.01178
  [hep-ph]} \BibitemShut {NoStop}%
%%CITATION = ARXIV:1512.01178;%%
\bibitem [{\citenamefont {Gleisberg}\ \emph {et~al.}(2009)\citenamefont
  {Gleisberg}, \citenamefont {Hoeche}, \citenamefont {Krauss}, \citenamefont
  {Schonherr}, \citenamefont {Schumann}, \citenamefont {Siegert},\ and\
  \citenamefont {Winter}}]{Gleisberg:2008ta}%
  \BibitemOpen
  \bibfield  {author} {\bibinfo {author} {\bibfnamefont {T.}~\bibnamefont
  {Gleisberg}}, \bibinfo {author} {\bibfnamefont {S.}~\bibnamefont {Hoeche}},
  \bibinfo {author} {\bibfnamefont {F.}~\bibnamefont {Krauss}}, \bibinfo
  {author} {\bibfnamefont {M.}~\bibnamefont {Schonherr}}, \bibinfo {author}
  {\bibfnamefont {S.}~\bibnamefont {Schumann}}, \bibinfo {author}
  {\bibfnamefont {F.}~\bibnamefont {Siegert}}, \ and\ \bibinfo {author}
  {\bibfnamefont {J.}~\bibnamefont {Winter}},\ }\href {\doibase
  10.1088/1126-6708/2009/02/007} {\bibfield  {journal} {\bibinfo  {journal}
  {JHEP}\ }\textbf {\bibinfo {volume} {02}},\ \bibinfo {pages} {007} (\bibinfo
  {year} {2009})},\ \Eprint {http://arxiv.org/abs/0811.4622} {arXiv:0811.4622
  [hep-ph]} \BibitemShut {NoStop}%
%%CITATION = ARXIV:0811.4622;%%
\bibitem [{\citenamefont {Bothmann}\ \emph {et~al.}(2019)\citenamefont
  {Bothmann} \emph {et~al.}}]{Bothmann:2019yzt}%
  \BibitemOpen
  \bibfield  {author} {\bibinfo {author} {\bibfnamefont {E.}~\bibnamefont
  {Bothmann}} \emph {et~al.},\ }\href {\doibase 10.21468/SciPostPhys.7.3.034}
  {\bibfield  {journal} {\bibinfo  {journal} {SciPost Phys.}\ }\textbf
  {\bibinfo {volume} {7}},\ \bibinfo {pages} {034} (\bibinfo {year} {2019})},\
  \Eprint {http://arxiv.org/abs/1905.09127} {arXiv:1905.09127 [hep-ph]}
  \BibitemShut {NoStop}%
%%CITATION = ARXIV:1905.09127;%%
\end{thebibliography}%

\end{document}